\documentclass{article}

\usepackage{url,amsmath,amsthm,amssymb}
\usepackage{natbib}

\usepackage{caption,subfig}
\usepackage{tikz}
\usetikzlibrary{shapes,arrows,positioning}

\let\hat\widehat

\begin{document}


\begin{center}
\textsf{\textbf{\Large Topological Data Analysis}}\\
\textsf{\textbf{Larry Wasserman}}\\
\textsf{\textbf{Department of Statistics/Carnegie Mellon University}}\\
\textsf{\textbf{Pittsburgh, USA, 15217; email: larry@stat.cmu.edu}}
\end{center}

\begin{abstract}
Topological Data Analysis (TDA)
can broadly be described as a collection of data analysis methods
that find structure in data.
This includes: clustering, manifold estimation,
nonlinear dimension reduction, mode estimation, ridge estimation and
persistent homology.
This paper reviews some of these methods.
\end{abstract}

\tableofcontents

\section{INTRODUCTION}

Topological Data Analysis (TDA)
refers to statistical methods that 
find structure in data.
As the name suggests, these methods
make use of
topological ideas.
Often, the term TDA is used narrowly to describe a particular method called
{\em persistent homology}
(discussed in Section \ref{section::persistence}).
In this review, I take a broader perspective:
I use the term TDA to refer to a large class
of data analysis method that uses notions of shape and connectivity.
The advantage of taking this broader definition of TDA is that it 
provides more context for recently developed methods.
The disadvantage is that my review must necessarily be incomplete.
In particular, I omit any reference to classical notions of shape
such as shape manifolds
\citep{kendall1984shape,
patrangenaru2015nonparametric} and related ideas.

Clustering is
the simplest example of TDA.
Clustering
is a huge topic and I will only
discuss {\em density clustering} since this connects clustering
to other methods in TDA.
I will also selectively review
aspects of manifold estimation (also called ``manifold learning''),
nonlinear dimension reduction, mode and ridge estimation and
persistent homology.

In my view,
the main purpose of TDA is to help
the data analyst summarize and visualize complex datasets.
Whether or not TDA can be used to make scientific discoveries
is still unclear.
There is another field that
deals with the topological and geometric structure of data:
computational geometry.
The main difference is that in TDA we treat the data as random points
whereas in 
computational geometry the data are usually seen as fixed.

Throughout this paper, we assume that we observe a sample
\begin{equation}
X_1,\ldots, X_n \sim P
\end{equation}
where the distribution $P$ is supported
on some set ${\cal X}\subset\mathbb{R}^d$.
Some of the technical results cited
require either that $P$ have sufficiently thin tails
or that ${\cal X}$ be compact.

\vspace{.2cm}

{\bf Software}: many of the methods in this paper
are implemented in the R package
{\tt TDA} available at
\url{https://cran.r-project.org/web/packages/TDA/index.html}.
A tutorial on the package can be found in
\cite{fasy2014introduction}.

\section{DENSITY CLUSTERS}
\label{sec::density}

Clustering is perhaps the oldest and simplest
version of TDA.
The connection between clustering and topology is clearest
if we focus on density-based methods for clustering.

\subsection{Level Set Clusters}

Let $X_1,\ldots, X_n$ be a random sample from a distribution $P$
with density $p$
where $X_i \in {\cal X}\subset \mathbb{R}^d$.
Density clusters are sets with high density.
\cite{hartigan1975clustering,hartigan1981consistency}
formalized this as follows.
For any $t\geq 0$
define the {\em upper level set}
\begin{equation}
L_t = \Bigl\{x:\ p(x) > t \Bigr\}.
\end{equation}
The density clusters at level $t$, denoted by ${\cal C}_t$,
are the connected components of $L_t$.
The set of all density clusters is
\begin{equation}
{\cal C} = \bigcup_{t\geq 0} {\cal C}_t.
\end{equation}
The leftmost plot in Figure \ref{fig::levelsetsandtree} shows a density function.
The middle plot 
shows the level set clusters corresponding to one particular value of $t$.

The estimated upper level set
is
\begin{equation}
\hat L_t = \Bigl\{x:\ \hat p(x) > t \Bigr\}
\end{equation}
where $\hat p$ is any density estimator.
A common choice is the kernel density estimator
\begin{equation}
\hat p_h(x) = \frac{1}{n}\sum_{i=1}^n \frac{1}{h^d}
K\left( \frac{||x-X_i||}{h}\right)
\end{equation}
where
$h>0$ is the bandwidth and $K$ is the kernel.
The theoretical properties of the estimator $\hat L_t$
are discussed, for example, in 
\cite{cadre2006kernel} and \cite{rinaldo2010generalized}.
In particular,
\cite{cadre2006kernel} shows, under regularity conditions and appropriate $h$,
that $\mu(\hat L_t \Delta L_t)=O_P(1/\sqrt{n h^d})$
where $\mu$ is Lebesgue measure and
$A\Delta B$ is the set difference between two sets $A$ and $B$.

To find the clusters, we need to get the connected components
of $\hat L_t$.
Let
$I_t = \{i:\ \hat p_h(X_i) > t\}$.
Create a graph whose nodes correspond to
$(X_i:\ i\in I_t)$.
Put an edge between two nodes $X_i$ and $X_j$ if
$||X_i - X_j|| \leq \epsilon$
where $\epsilon>0$ is a tuning parameter.
(In practice $\epsilon = 2h$ often seems to work well.)
The connected conponenets 
$\hat C_1, \hat C_2, \ldots$
of the graph
estimate the clusters at level $t$.
The number of connected components is denoted by
$\beta_0$ which is the zeroth-order Betti number.
This is discussed in more detail in Section \ref{section::homology}.

Related to level sets is the concept of excess mass.
Given a class of sets
${\cal C}$, 
the {\em excess mass functional} is defined to be
\begin{equation}
E(t) = \sup\{ P(C)-t \mu(C):\ C \in {\cal C}\}
\end{equation}
and any set $C\in {\cal C}$ such that
$P(C)-t\mu(C) = E(t)$
is called a generalized $t$-cluster.
If ${\cal C}$ is taken to be all measurable sets
and the density is bounded and continous, then
the upper level set $L_t$ is the unique $t$-cluster.
The excess
mass functional is 
studied in 
\cite{polonik1995measuring, muller1991excess}.

One question that arises
in the use of level set clustering is: how do we choose $t$?
One possibility is to choose $t$ 
to cover some prescribed fraction $1-\beta$ of the total mass;
thus we choose $t$ to satisfy
$\int_{\hat L_t} \hat p(s) ds = 1-\beta$.
Another idea is to look at clusters
at all levels $t$.
This leads us to the idea of density trees.

\subsection{Density Trees}
\label{section::trees}

\begin{figure}
\begin{center}
\includegraphics[width=5in,height=2in]{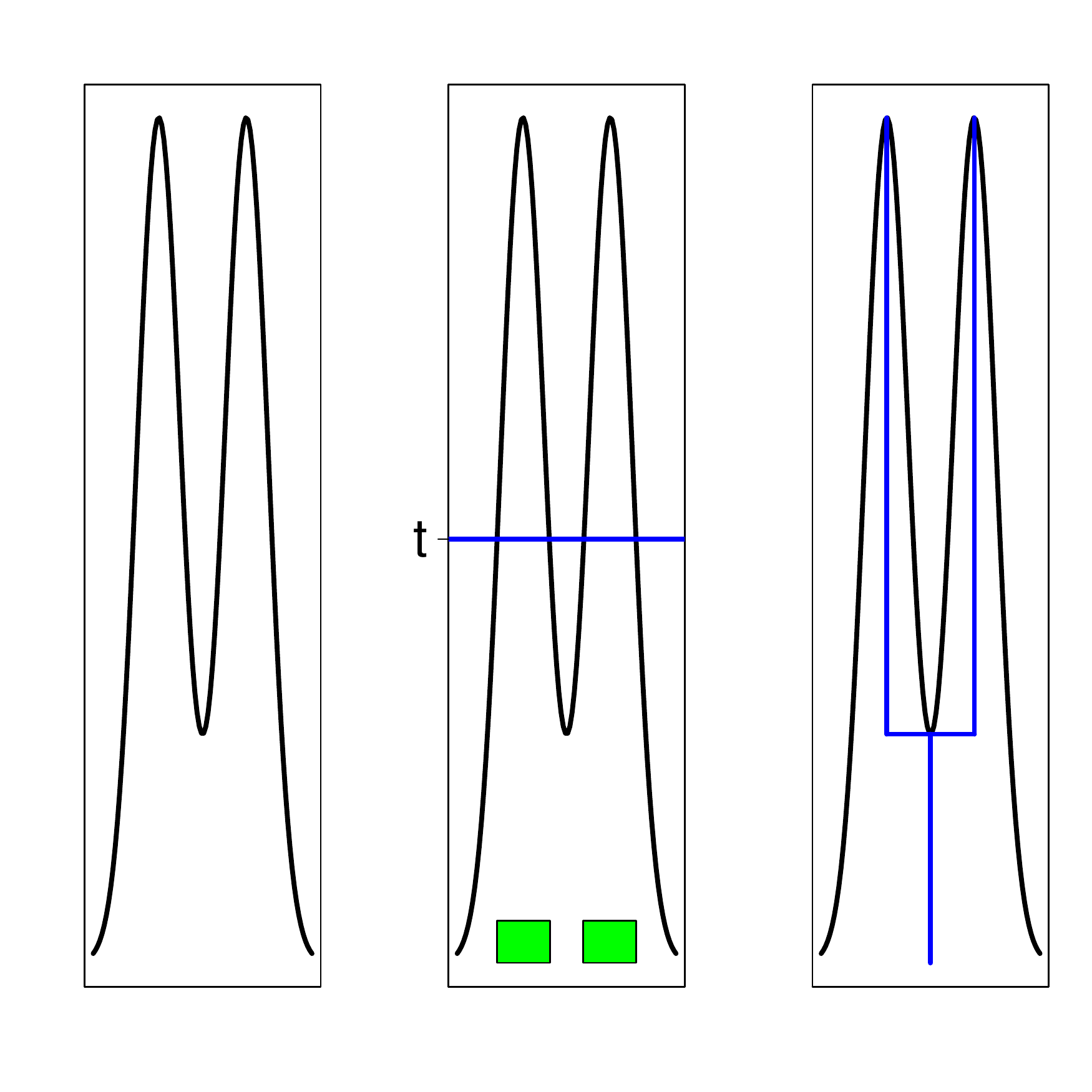}
\end{center}
\caption{Left: a density function $p$.
Middle: density clusters corresponding to $L_t=\{x:\ p(x) > t\}$.
Right: the density tree corresponding to $p$ is shown under
the density.
The leaves of the tree correspond to modes. The branches correspond to
connected components of the level sets.}
\label{fig::levelsetsandtree}
\end{figure}

The set of all density clusters ${\cal C}$
has a tree structure:
if $A,B \in {\cal C}$ then either
$A\subset B$ or $B\subset A$ or
$A\bigcap B = \emptyset$.
For this reason, we can visually represent a density and its clusters
as a tree which we denote by $T_p$ or $T(p)$.
Note that $T_p$ is technically a collection of level sets,
but it can be represented as a two-dimensional tree as in
the right-most plot in Figure \ref{fig::levelsetsandtree}.
The tree, shown under the density function,
shows the number of level sets and
shows when level sets merge.
For example, if we cut across at some level $t$, then the number of braches of the tree
corresponds to the number of connected components of the level set.
The leaves of the tree correspond to the modes of the density.

The tree is called a {\em density tree} or {\em cluster tree}.
This tree provides a convenient, two-dimensional visualization of a density
regardless of the dimension $d$ of the space in which the data lie.

\begin{figure}
\begin{center}
\begin{tabular}{ccc}
\includegraphics[scale=.2]{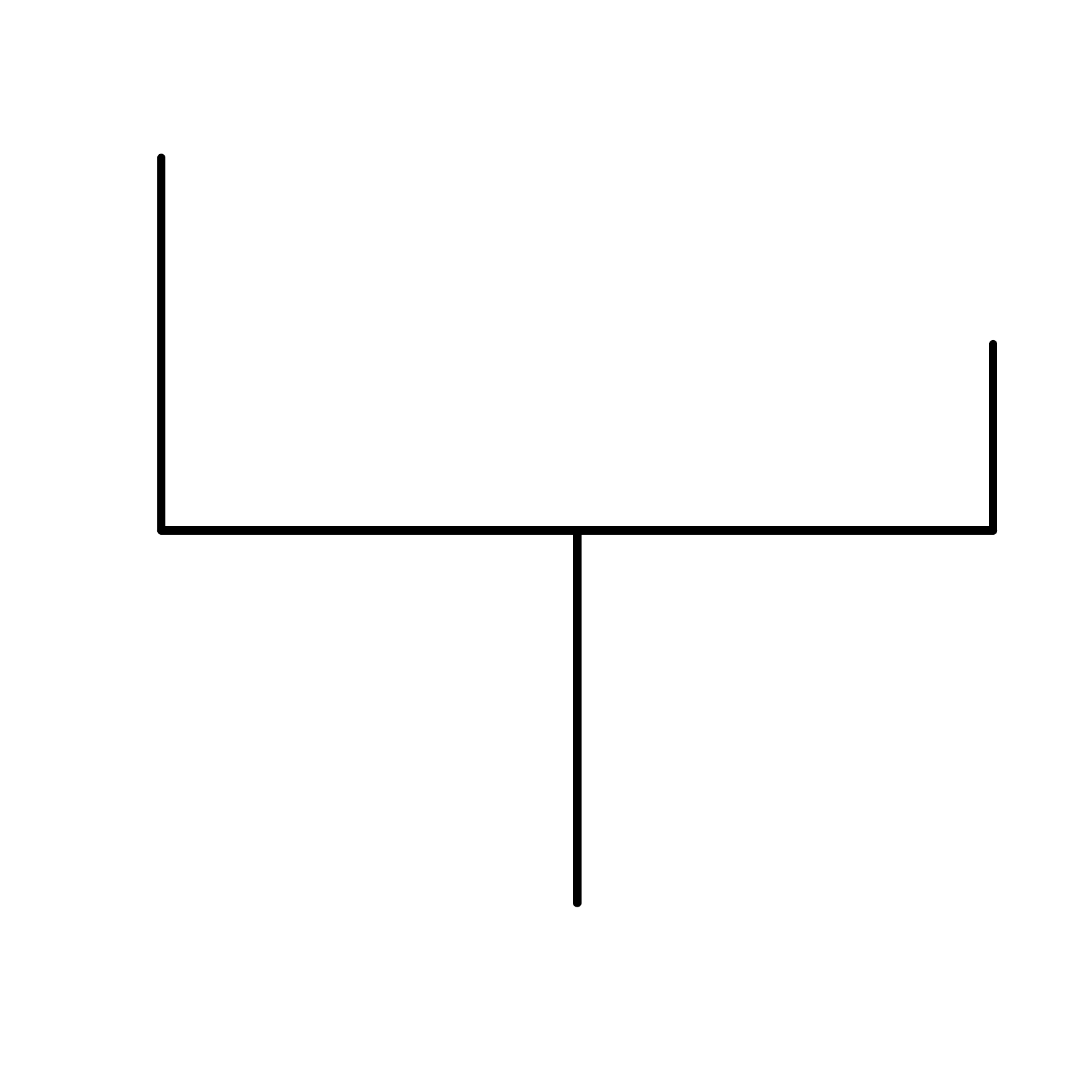} & 
\includegraphics[scale=.2]{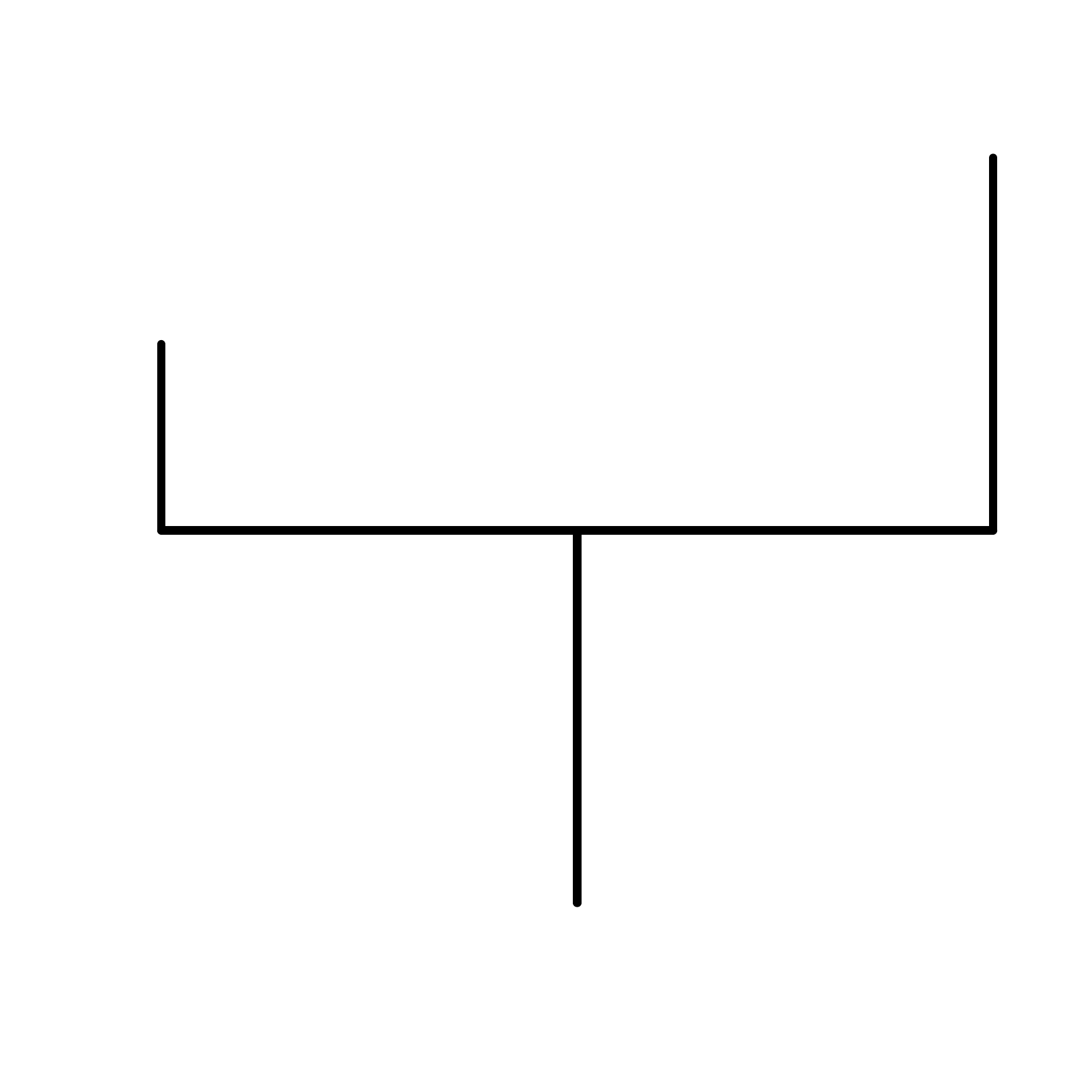} & 
\includegraphics[scale=.2]{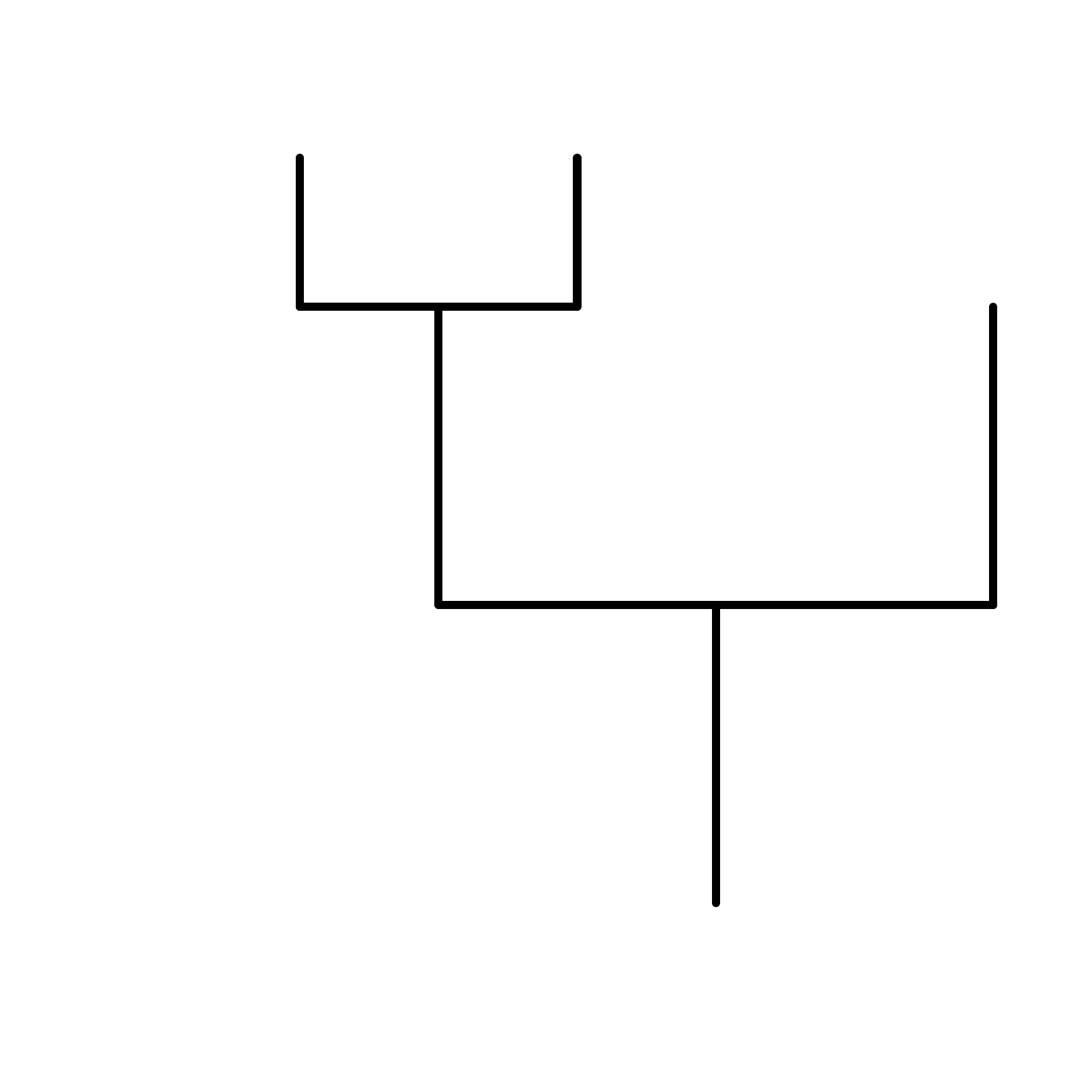}
\end{tabular}
\end{center}
\caption{The first and second density trees are homeomorphic; there exists
a bi-continous map from one tree to the other. 
The third tree is not homeomorphic to the other two.
Thus the first two trees represent densities with the same shape.}
\label{fig::tree-homeomorphism}
\end{figure}

Two density trees have the same ``shape''
if their tree structure is the same.
\cite{chen2016statistical} make this precise as follows.
For a given tree $T_p$ define a distance on the tree by
$$
d_{T_p}(x,y) = |p(x) + p(y) - 2m_p(x,y)|
$$
where
$$
m_p(x,y) = \sup\{t:\ {\rm there\ exists\ }C\in {\cal C}_t\ {\rm such\ that\ }x,y\in C\}
$$
is called the 
merge height 
\citep{eldridge2015beyond}.
For any two clusters $C_{1}, C_2 \in T_p$, we first
define $\lambda_1 = \sup \{t: C_1 \in {\cal C}_t\}$, and
$\lambda_2$ analogously. We then define the tree distance function
on $T_p$ by
\begin{equation}
d_{T_{p}}(C_{1},C_{2})=\lambda_{1}+\lambda_{2}-2m_{p}(C_{1},C_{2})
\end{equation}
where
$$
m_p(C_1,C_2)=\sup\{\lambda\in\mathbb{R}:\ 
{\rm there\ exists\ }C\in T_p\ {\rm such\ that\ } C_1,C_2\subset C\}.
$$
Now $d_{T_{p}}$ defines a distance on the tree and
it induces a topology on $T_p$.
Given two densities $p$ and $q$,
we say $T_{p}$ is homeomorphic to $T_{q}$, written 
$T_{p}\cong T_{q}$, 
if there exists a bicontinuous map from $T_p$ to $T_q$.
This means that $T_p$ and $T_q$ have the same shape.
In other words, they have the same tree structure.
An example is shown in Figure \ref{fig::tree-homeomorphism}.

The density tree can be estimated by plugging in any density estimator.
The estimated tree is denoted by $\hat T$ --- usually based on a kernel
density estimator $\hat p_h$
which provides a nice visualization of the cluster structure
of the data.
Another choice of estimator
is the $k$-nearest neighbor estimator
as in \cite{chaudhuri2010rates}.

To estimate the shape of the density tree, it is not necessary to let the bandwidth $h$ go to 0 as $n$ increases.
Let $p_h (x) = \mathbb{E}[\hat p_h(x)]$ be the mean of the estimator.
It can be shown 
that, under weak conditions,
there exists $h_0 > 0$ such that, for all $0 < h < h_0$,
$T(p_h)\cong T(p)$.
This means that it suffices to estimate $T_{p_h}$ for any small $h>0$.
It is not necessary to let $h\to 0$.
This has important pratical implications since
$T_{p_h}$ can be estimated at the rate $O_P(n^{-1/2})$ independent of the dimensions $d$.
Compare this to estimating $p$ in the $L_2$ loss; the best rate
under standard smoothness conditions is $O_P(n^{-2/(4+d)})$ which is slow for large dimensions $d$.
The key point is: estimating the cluster structure is easier than estimating the density itself.
In other words, you can estimate $p$ poorly but still get the shape of the tree correct.
See \cite{chen2016statistical} for more details.

\begin{figure}
\mbox{\includegraphics[height=2.5 in]{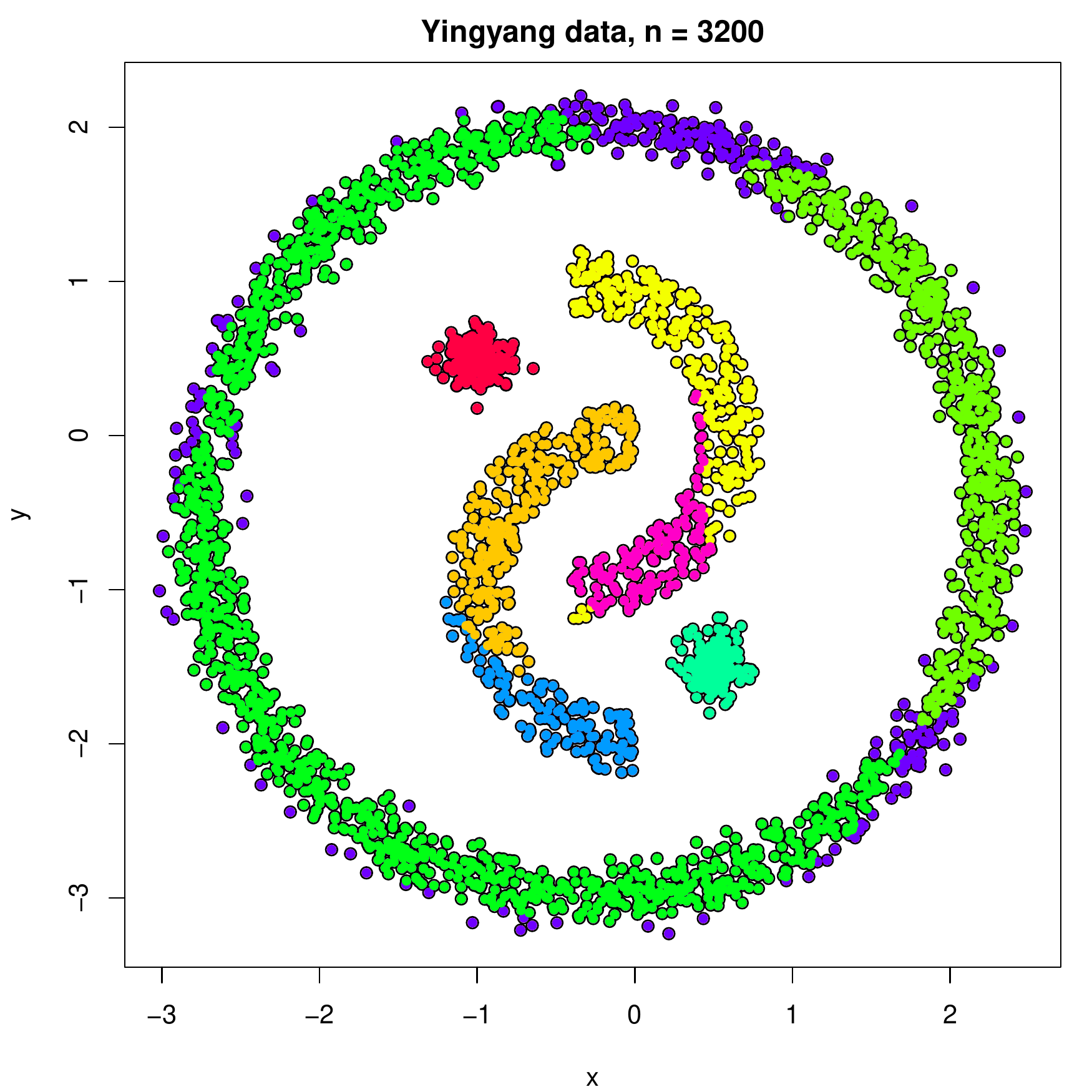}
\includegraphics[height=2.5 in]{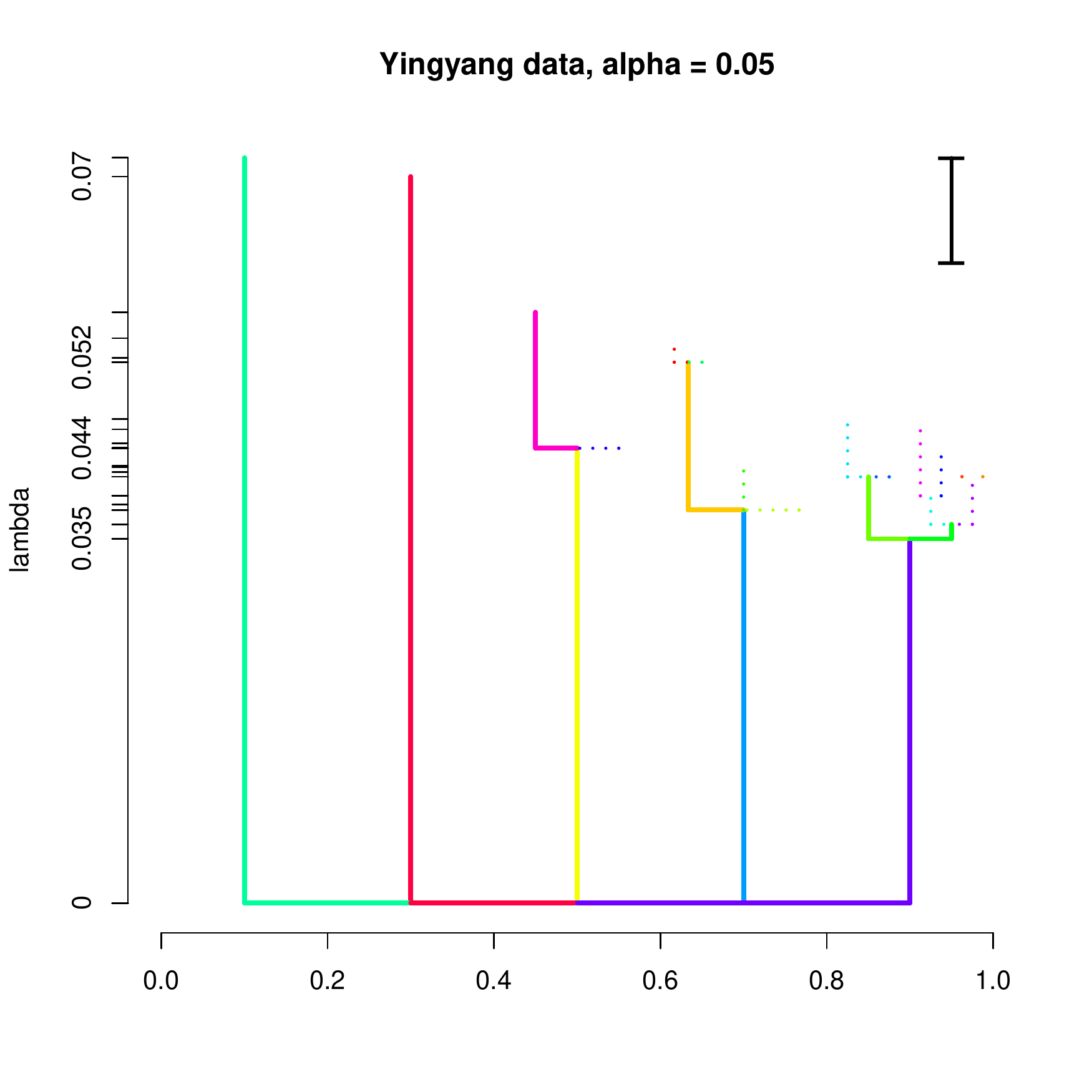}}
\vspace{.1cm}
\caption{Example from \cite{chen2016statistical}.
Left: the data.
Right: the tree.
The solid lines are the pruned trees; 
The dashed lines are leaves and branches that have been pruned away
because they are smaller than the bootstrap
significance level
$2\hat{t}_\alpha$ (indicated in the top right corner).}
\label{fig::tree-example}
\end{figure}

The bootstrap can be used to get confidence sets for the
density tree \citep{chen2016statistical}.
Let $P_n$ be the empirical measure that puts mass $1/n$ at each data point.
Draw an iid sample $X_1^*,\ldots, X_n^* \sim P_n$
and compute the density estimator
$\hat p_h^*$.
Repeat this process $B$ times 
to get density estimates
$\hat p_h^{*(1)},\ldots, \hat p_h^{*(B)}$
and define
$$
\hat F_n(t) =\frac{1}{B}\sum_{j=1}^B 
I(\sqrt{n} ||\hat p_h^{*(j)}-\hat p_h||_\infty >t)
$$
where $I$ is the indicator function.
For large $B$, $\hat F_n$ approximates
$$
F_n(t) = P( \sqrt{n} ||\hat p_h-  p_h||_\infty >t).
$$
Let
$\hat t_\alpha = \hat F_n^{-1}(1-\alpha)$
which approximates
$t_\alpha = F_n^{-1}(1-\alpha)$.
Then
$$
\lim_{n\to\infty}P( T(p_h)\in {\cal T})= 1-\alpha
$$
where
$$
{\cal T} = \Biggl\{ T(p):\ ||p-\hat p_h||_\infty \leq \frac{\hat t_\alpha}{\sqrt{n}}\Biggr\}.
$$
Thus,
${\cal T}$ is an asymptotic confidence set for the tree.
The critical value $\hat t_\alpha$ can be used to
prune non-significant leaves and branches from $\hat T$;
see Figure \ref{fig::tree-example}.

A density tree is
{\em Hartigan consistent}
if, with probability tending to 1, 
the correct cluster strucrure is recovered.
Generally, density trees based on consistent density estimators
will be Hartigan consistent.
For more on Hartigan consistency, see
\cite{chaudhuri2010rates, eldridge2015beyond, balakrishnan2013cluster}.

\subsection{Mode Clustering and Morse Theory}
\label{sec::modes}

Another density clustering method
is {\em mode clustering}
\citep{chacon2015population, chacon2013data, chacon,
li2007nonparametric, comaniciu2002mean, arias2015estimation, cheng1995mean}.
The idea is to find modes of the density and then define
clusters as the basins of attraction of the modes.
A point $m$ is a (local) mode
if there exists an open neighborhood $N$ of $x$ such that
$p(x) > p(y)$ for every $y\in N$ such that $y\neq x$.
Suppose that $p$ has $k$ local modes
${\cal M} =\{m_1,\ldots, m_k\}$.
Assume that $p$ has gradient $g$ and Hessian $H$.

A point $x$ is a {\em critical point} if
$g(x) = (0,\ldots,0)^T$.
The function $p$ is a 
{\em Morse function} if the Hessian is non-degenerate at each
critical point
\citep{milnor2016morse}.
We will assume that $p$ is Morse.
In this case, $m$ is a local mode
if and only if
$g(m) = (0,\ldots, 0)^T$ and
$\lambda_1(H(m)) < 0$
where
$\lambda_1(A)$ denotes the largest eigenvalue of the matrix $A$.

\begin{figure}
\begin{center}
\mbox{
\includegraphics[scale=.5]{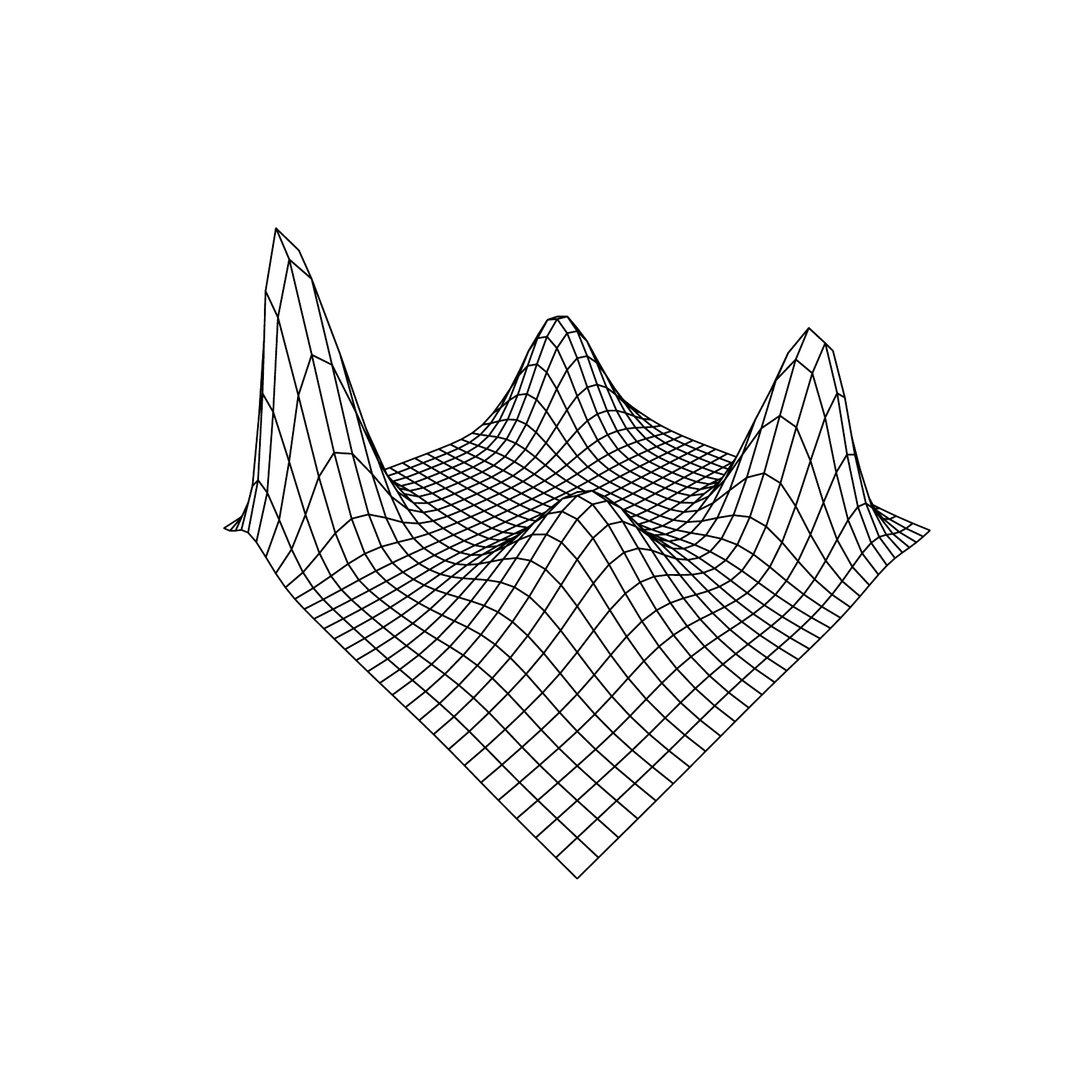}%
\includegraphics[scale=.4]{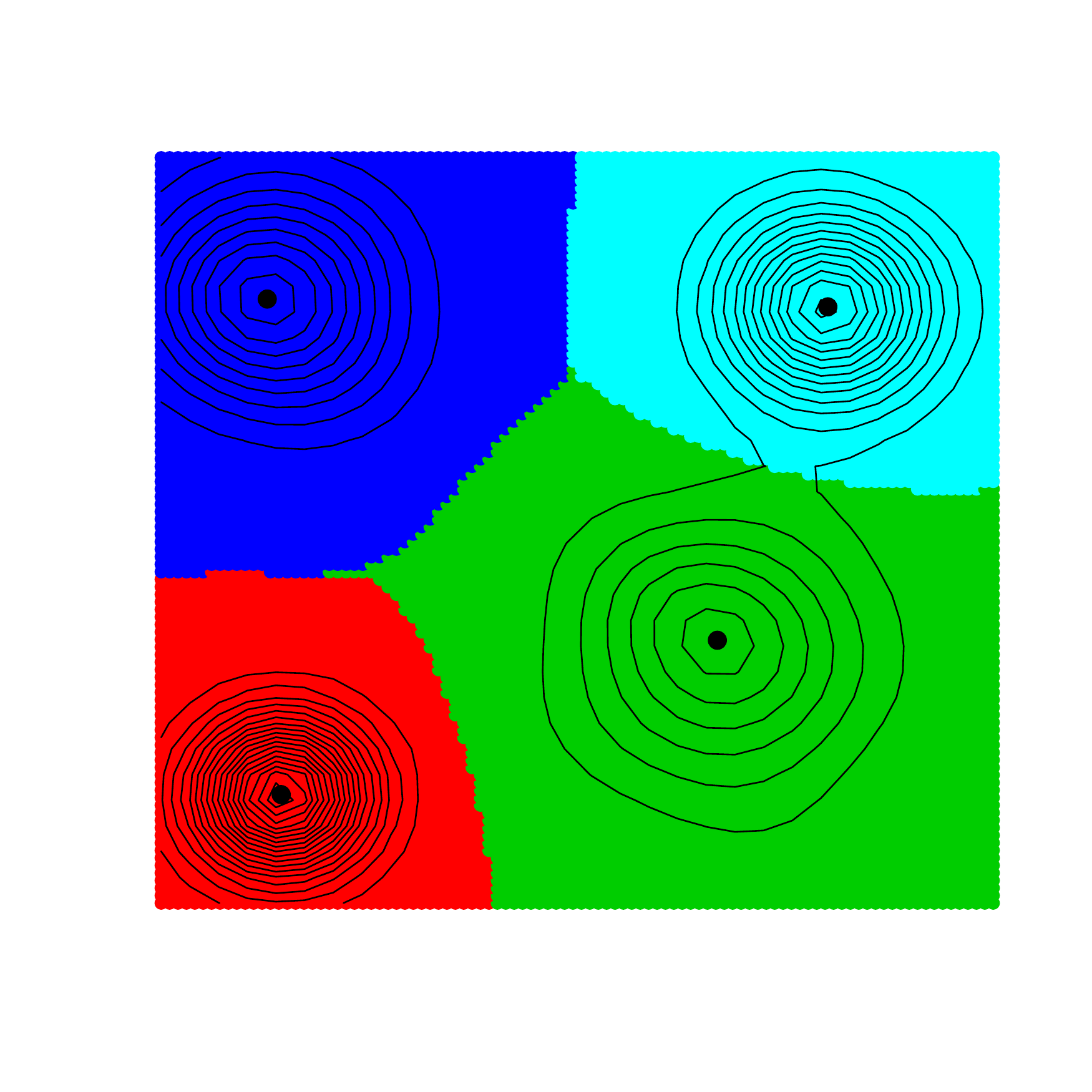}}
\end{center}
\caption{Left: a density with four modes. Right: the partition 
(basins of attraction) of the space induced
by the modes. These are the population clusters.}
\label{fig::mode-fig}
\end{figure}

Now let $x$ be an arbitrary point.
If we follow the steepest ascent path starting at $x$,
we will eventually end up at one of the modes.\footnote{This is true for all $x$
except on a set of Lebesgue measure 0.}
Thus, each point $x$ in the sample space is assigned to a mode $m_j$.
We say that $m_j$ is the {\em destination} of $x$ which is written
$$
{\rm dest}(x) = m_j.
$$
The path $\pi_x: \mathbb{R}\to\mathbb{R}^d$
that leads from $x$ to a mode is defined by the
differential equation
$$
\pi_x'(t) = \nabla p(\pi_x(t)),\ \ \ \pi_x(0) =x.
$$
The set of points assigned to mode $m_j$
is called the
{\em basin of attraction} of $m_j$ and is denoted by $C_j$.
The sets $C_1,\ldots, C_k$
are the population clusters.
The left plot in Figure \ref{fig::mode-fig} shows a bivariate density
with four modes.
The right plot shows the partition induced by the modes.

\begin{figure}
\begin{center}
\includegraphics[width=4in,height=2in]{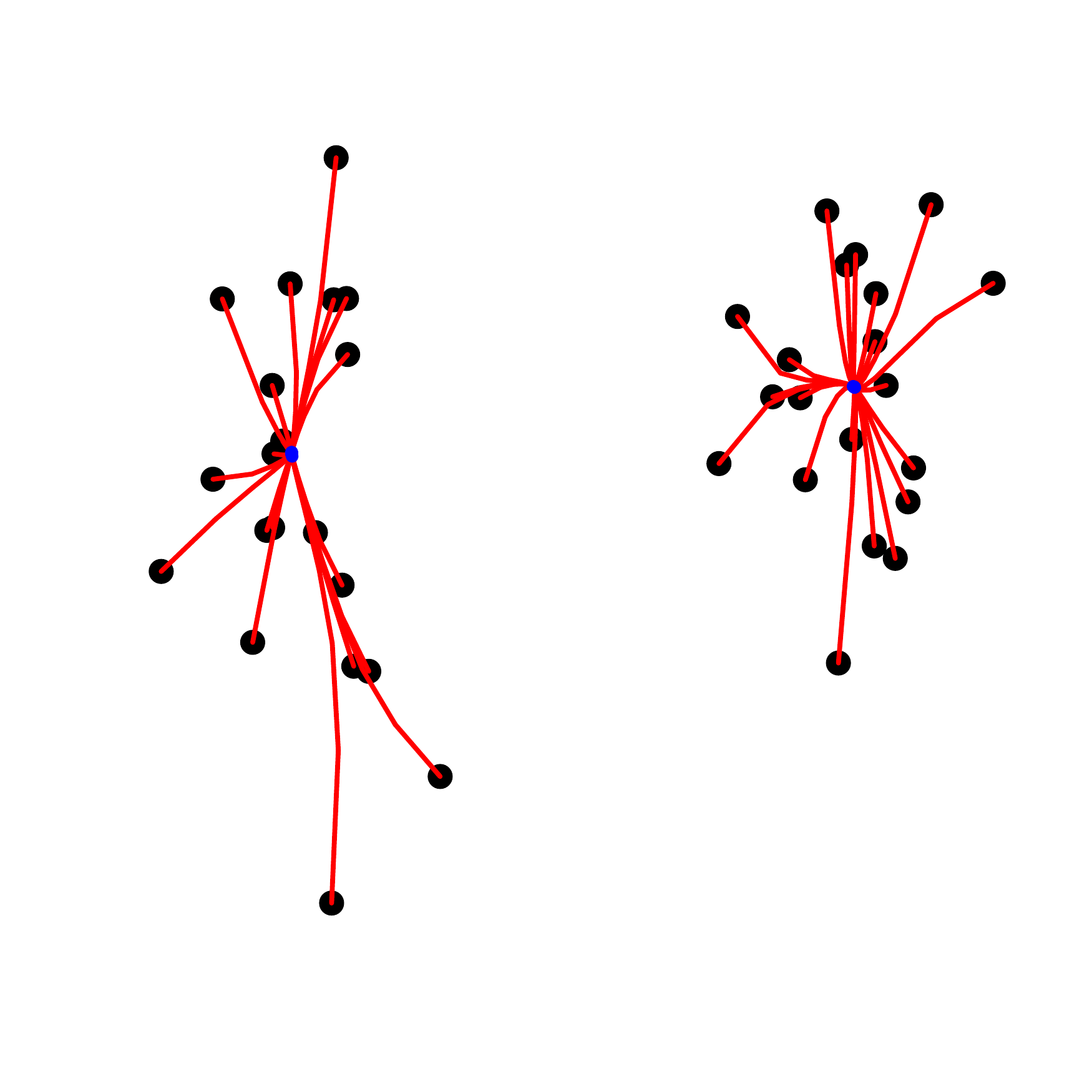}
\end{center}
\vspace{.2cm}
\vspace{-.1in}
\caption{The mean shift algorithm.
The data are represented by the black dots.
The modes of the density estimate are the two blue dots.
The red curves show the mean shift paths; each data point moves along its path towards a mode
as we iterate the algorithm.}
\label{fig::simplemeanshift}
\end{figure}

To estimate the
clusters, we find the modes
$\hat {\cal M} =\{\hat m_1,\ldots, \hat m_r\}$
of the density estimate.
A simple algorithm called the
{\em mean shift algorithm}
\citep{cheng1995mean, comaniciu2002mean}
can be used to find the modes and to
find the destiation of a any point $x$.
For any given $x$, we define the iteration
$$
x^{(j+1)} = \frac{\sum_i X_i K\left(\frac{||x^{(j)}-X_i||}{h}\right)}
{\sum_i K\left(\frac{||x^{(j)}-X_i||}{h}\right)}.
$$
See Figure \ref{fig::simplemeanshift}.
The convergence of this algorithm is studied in
\cite{arias2015estimation}.

It can be shown under suitable regularity conditions that
the modes of the kernel density estimate
are consistent estimates modes of the true density;
see \cite{genovese2016non}.
Once again, however, it is not necessary to estimate the density well
to estimate the mode clusters well.
Specifically, define
$$
c(x,y) =
\begin{cases}
1 & {\rm if\ } {\rm dest}(x) = {\rm dest}(y)\\
0 & {\rm if\ } {\rm dest}(x) \neq {\rm dest}(y).
\end{cases}
$$
Thus, $c(x,y)=1$ if $x$ and $y$ are in the same cluster.
Similarly, the estimated clusters define a function $\hat c$.
Let $C_1,\ldots, C_k$ be the model clusters.
Let $t_1,\ldots, t_k$ be constants and let
$C_j(t_j) = \{x\in C_j:\ p(x)> t_j\}$.
The sets $C_1(t_1),\ldots, C(t_k)$ are called
{\em cluster cores}.
These are the high density points within the clusters.
Let ${\rm Core} = \{X_i:\ X_i\in \bigcup_j C_j(t_j)\}$
be the data points in the cluster cores.
\cite{azizyan2015risk} show that,
if $t_1,\ldots, t_k$ are sufficiently large,
then
$$
\mathbb{P}(\hat c(X_j,X_k) \neq c(X_j,X_k)\ {\rm for\ any\ }X_j,X_k\in {\rm Core})\leq
e^{-nb}
$$
for some $b>0$, independent of the dimension.
This means that high density points can be accurately clustered using mode
clustering.

\section{LOW DIMENSIONAL SUBSETS}

Sometimes the distribution $P$ is supported on
a set $S$ of dimension $r$ with $r < d$.
(Recall that $X_i$ has dimension $d$.)
The set $S$ might be of scientific interest 
and it is also useful for dimension reduction.
Sometimes the support of $P$ is $d$-dimensional but we are interested in finding a set $S$
of dimension $r<d$ which has a high concentration of mass.

\begin{figure}
\begin{center}
\includegraphics[scale=.4]{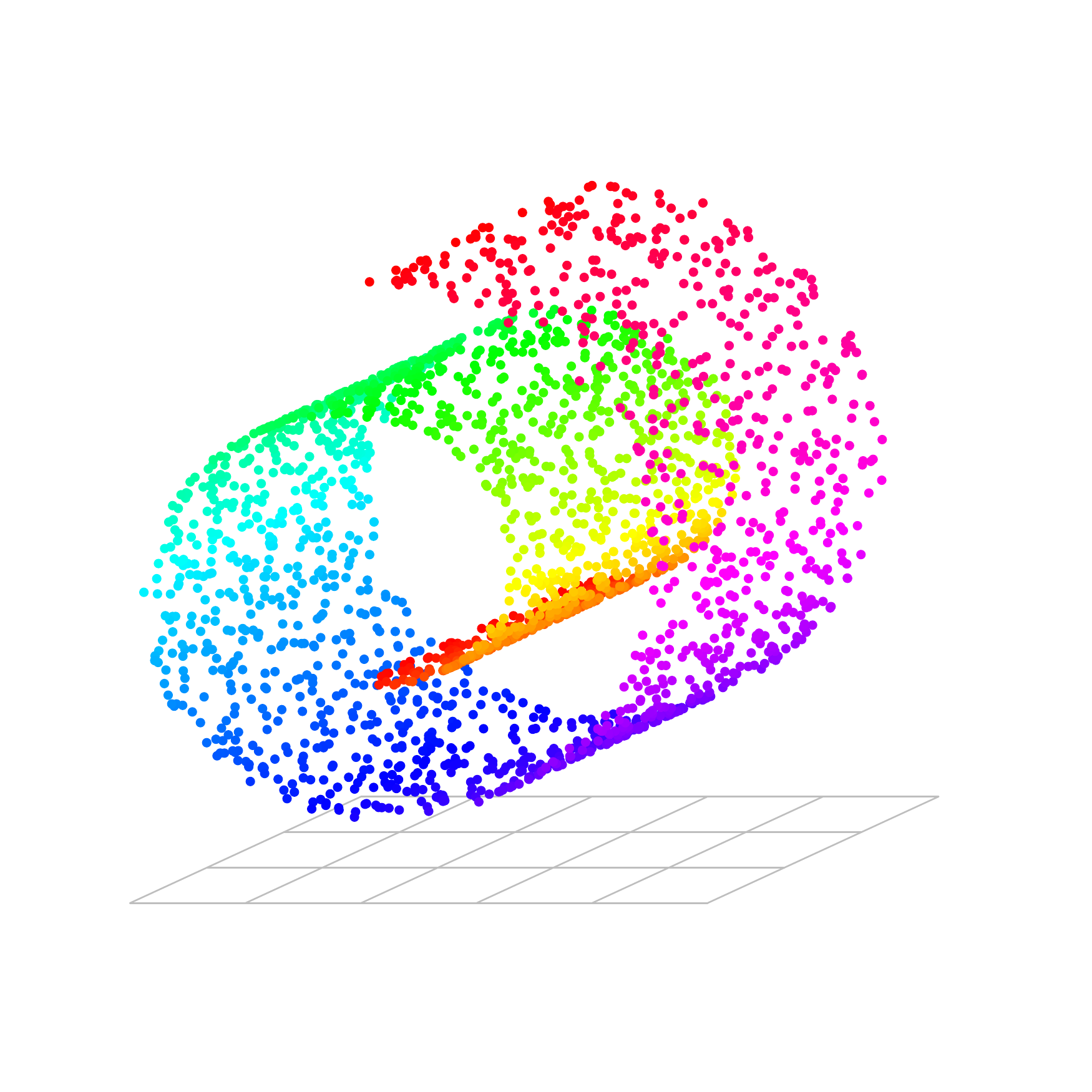}
\end{center}
\caption{The swissroll dataset. The ambient dimension is $d=3$ but the data
are supported on a set $S$ of dimension $r=2$.}
\label{fig::swiss}
\end{figure}

\begin{figure}
\begin{center}
\includegraphics[scale=.5]{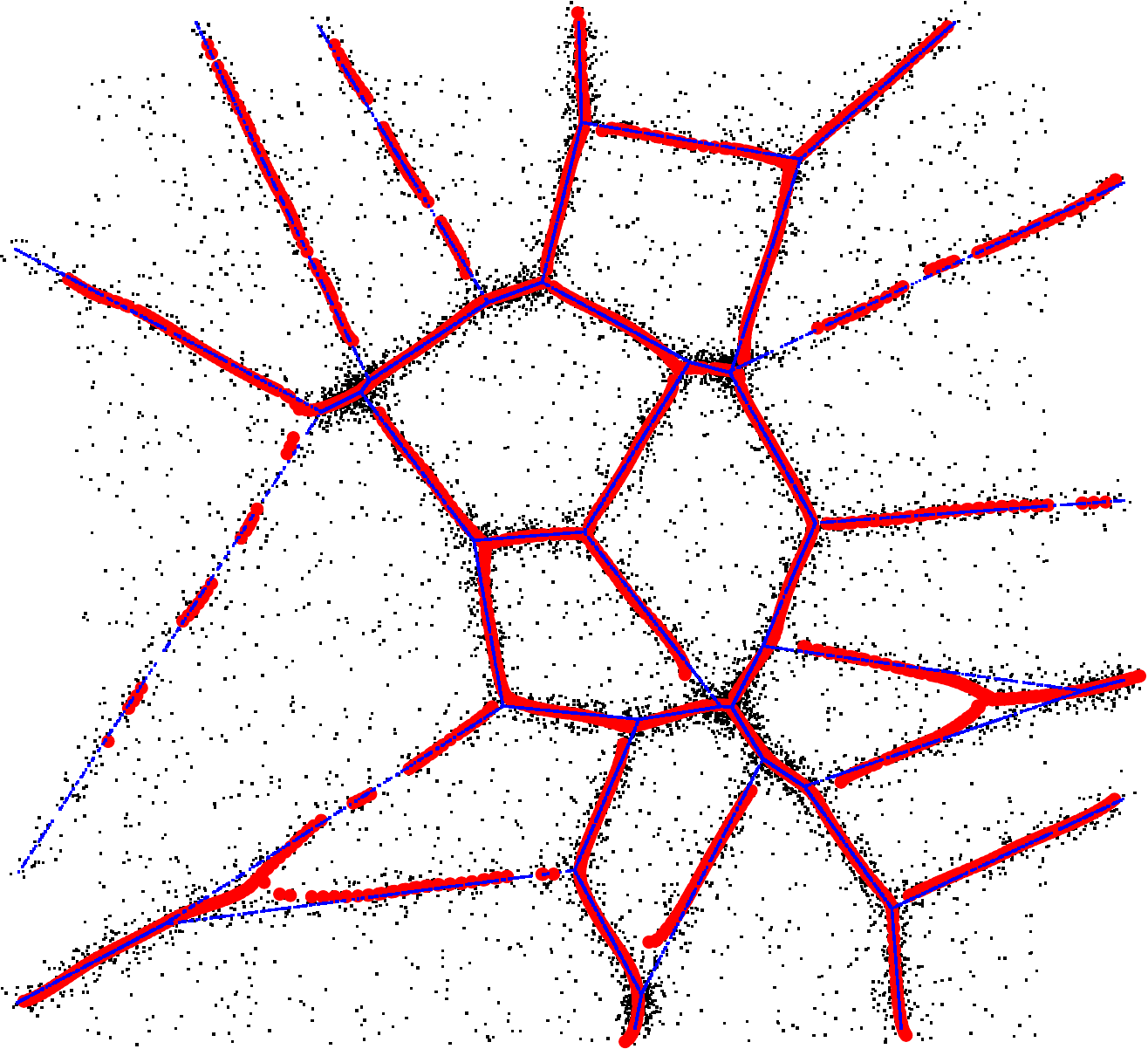}
\end{center}
\caption{These data are two-dimensional but there is a set $S$ of dimension $r=1$
with a high concentration of data.
The red lines show an estmate of $S$ using the methods in Section 
\ref{section::ridges}.}
\label{fig::web}
\end{figure}

Figure \ref{fig::swiss} shows an example known as the Swiss-roll dataset.
Here, the ambient dimension is $d=3$ but the support
of the distribution $S$ is a manifold of instrinsic dimension $r=2$.
Figure \ref{fig::web} shows a more complex example.
Here, $d=2$ but clearly there is a $r=1$ intrinsic dimensional subset $S$
with a high concentration of data.
(This dataset mimics what we often see in some datasets from astrophysics.)
The set $S$ is quite complex and is not a smooth manifold.
The red lines show an estimate of $S$ based on the techniques
described in Section \ref{section::ridges}.

\subsection{Manifolds}

In the simplest case,
the set $S$ is a smooth, compact submanifold
of dimension $r$.
The term
{\em manifold learning}
can refer either to methods for estimating the set $S$
or to dimension reduction methods that
assume that the data are on
(or near) a manifold.
Principal component analysis
can be thought of as a special case of manifold learning
in which the data are assumed to lie near an affice subspace.

As a motivating example, consider
images of a person's face as the person moves their head.
Each image can be regareded as a high-dimensional vector.
For example, a 16 by 16 image is a vector in
$\mathbb{R}^d$ where $d=16\times 16 = 256$.
However, the set of images will not fill up 
$\mathbb{R}^{256}$.
As the person moves their head,
these vectors are likely to trace out a surface of dimension $r=3$,
corresponding to the three degrees of freedom corresponding to the motion of the head.

\vspace{1cm}

{\bf Estimating $S$.}
An estimator of $S$ is 
$\hat S = \bigcup_{i=1}^n B(X_i,\epsilon_n)$
which was suggested (in a different context) by
\cite{devroye1980detection}.
The estimator $\hat S$ is $d$-dimensional
but it does converge to $S$ in the following sense
\citep{cuevas2009set, fasy2014confidence, niyogi2008finding,
cuevas2001cluster, chazal2014convergence}.
The {\em Hausdorff distance} $H(A,B)$ 
between two sets $A$ and $B$ is
\begin{equation}\label{eq::hausdorff}
H(A,B) = \inf\{\epsilon:\ A \subset B\oplus \epsilon\ \ {\rm and}\ \ B\subset A\oplus \epsilon\}
\end{equation}
where
$$
A\oplus \epsilon = \bigcup_{x\in A} B(x,\epsilon)
$$
and 
$B(x,\epsilon)$ denotes a ball of radius $\epsilon$ centered at $x$.
Suppose there exists $c>0$ such that, for every $x\in S$ and every small $\epsilon$,
$P(B(x,\epsilon))\geq c \epsilon^r$.
Further, assume that the number of balls of size $\epsilon$ required to cover $S$ is
$C(1/\epsilon)^{r}$.
These assumption mean that $S$ is $r$-dimensional
(and not too curved) and that
$P$ spreads its mass over all of $S$.
Then
$$
P(H(\hat S,S) > \epsilon) \leq
C r^{-d} e^{-n c \epsilon^d}.
$$
Hence, if we choose
$\epsilon_n\asymp (\log n/n)^{1/r}$
then
$$
H(\hat S,S) = O_P\left( \frac{\log n}{n}\right)^{1/r}
$$
where we recall that $H$ is the Hausdorff distance defined in equation (\ref{eq::hausdorff}).
However, better rates are possible under some conditions.
The difficulty of estimating $S$ as defined by minimax theory is given 
under various sets of assumptions, in
\cite{us::2010,singular::2011}.

It is unlikey that a sample will fall precisely on a submanifold $S$.
A more realistic model
is that we observe
$Y_1,\ldots, Y_n$ where
$Y_i = X_i + \epsilon_i$
where
$X_1,\ldots, X_n \sim G$ is a sample from a distribution $G$ supported on $S$
and $\epsilon_1,\ldots, \epsilon_n$ are a sample from a noise distribution
such as a Gaussian.
In this case,
\cite{singular::2011} showed that
estimating $S$ is hopeless; the minimax rate of convergence is
logarithmic.
However, it is possible to
estimate an $r$-dimensional, high density region $R$
that is close to $S$.
The set $R$ corresponds to a ridge in the density of $Y$;
see Section \ref{section::ridges}.

\vspace{1cm}

{\bf Estimating the Topology of a Manifold.}
Another problem is to find an estimate $\hat S$ of $S$ that
is topologically similar to $S$.
If, for example, $S$ is a three dimensional image,
such as in Figure \ref{fig::rabbit},
then requring $\hat S$ to be topologically similar ensures that
$\hat S$ ``looks like'' $S$ in some sense.
But what does 
``topologically similar'' mean?

Two sets $S$ and $T$ (equipped with topologies) are {\em homeomorphic}
if there exists a bi-continous map from $S$ to $T$.
\cite{markov} proved that, in general,
the question of whether two spaces are homeomorphic is undecidable
for dimension greater than 4.

Fortunately, it is possible to determine if two
spaces are {\em homologically equivalent}.
Homology is way of defining topological features
algebraically using group theory.
The zero-th order homology of a set corresponds to
its connected components.
The first order homology corresponds to one-dimensional holes (like a donut).
The second order homology corresponds to two-dimensional holes (like a soccer ball).
And so on.
If two sets are homeomorphic then they are homologically equivalent.
However, the reverse is not true.
This, homological equivalence is weaker than topological equivalence.

We'll discuss homology in more detail in Section \ref{section::homology}.
But here, we mention one of the first results about
topology and statistics due to 
\cite{niyogi2008finding}.
They showed that
$$
\hat S = \bigcup_{i=1}^n B(X_i,\epsilon)
$$
has the same homology as $S$
with high probability, as long as 
$S$ has positive reach and
$\epsilon$ is small relative to the reach.
The {\em reach} of $S$ is the largest real number $r$ such that
any point $x$ that is a distance less than $r$ from $S$, has a unique projection on $S$.
The result assumes the data are sampled from a distribution supported 
on the submanifold $S$.
Extensions that allow for noise
are given in
\cite{niyogi2011topological}.
An unsolved problem is to find a data-driven method for choosing the tuning
parameter $\epsilon$.
The assumption that $S$ has positive reach can be weakened:
\cite{chazal2009sampling} define a quantity called that $\mu$-reach
which is weaker than reach and they show that
topological reconstructions are possible using this weaker regularity assumption.

\vspace{1cm}

{\bf Dimension Reduction.}
There are many methods that leverage 
the fact the the data are supported on a low
dimensional set $S$ without explicitly producing an estimate
$\hat S$ that is close to $S$ in Hausdorff distance.
Examples include:
Isomap \citep{tenenbaum2000global, de1999extended},
Local Linear Embedding \citep{roweis2000nonlinear},
diffusion maps \citep{coifman2006diffusion},
Laplacian eigenmaps \citep{belkin2001laplacian}
and many others \citep{lee2007nonlinear}.
Here, I will give a very brief description of Isomap.

The first step in Isomap is to form a graph from the data.
For example, we connect two points $X_i$ and $X_j$
if $||X_i - X_j|| \leq \epsilon$
where $\epsilon$ is a tuning parameter.
Next we define the distance between two points 
as the shortest path between the two points
among all paths in the graph that connect them.
We now have a distance matrix $D$ where
$D_{ij}$ is the shortest path between $X_i$ and $X_j$.
The hope is that $D_{ij}$ approximates the geodesic distance between $X_i$ and $X_j$ on the manifold.
Finally, we use a standard dimension reduction method
such as multidimensional scaling
to embed the data in $\mathbb{R}^r$ while trying
to preserve the distances $D_{ij}$ as closely as possible.
For example, we find a map $\phi$ to minimize the distorting
$\sum_{i<j}[ D^2_{i,j} - ||\phi(X_i)-\phi(X_j)||^2]$.
The transformed data
$Z_i = \phi(X_i)$ now live in a lower dimensional space.
Thus we have used the fact that the data 
live on a manifold, to perform a dimension reduction.

\begin{figure}
\begin{center}
\includegraphics[width=5in,height=3in]{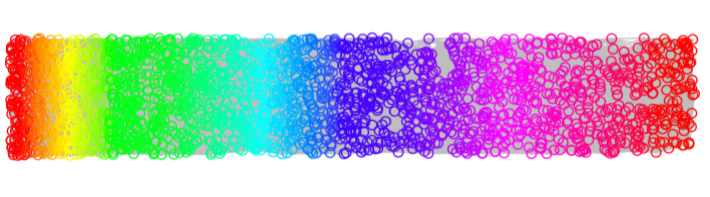}
\end{center}
\caption{After applying isomap
to the swissroll dataset with $\epsilon=5$ we recover
the undelrying two-dimensional structure.}
\label{fig::swiss2}
\end{figure}

Figure \ref{fig::swiss2} shows the result of applying isomap
to the swissroll data using $\epsilon=5$.
In this case we perfectly recover the underlying structure.
However, isomap is a fragile procedure.
It is very sensitive to outliers
and the choice of tuning parameters.
Other methods, such as diffusion maps
and ridge estimation, are more robust.

\subsection{Estimating Intrinsic Dimension}

Many manifold estimation methods
assume that the intrinsic dimension $r$ of the manifold is known.
In practice, we need to estimate the dimension.
There is a large literature on this problem.
Some examples include
\cite{little2011multiscale,
lombardi2011minimum,
hein2005intrinsic,
levina2004maximum,
kegl2002intrinsic,
costa2004geodesic}.
Minimax theory for dimension estimation is contained in
\cite{Koltchinskii} and \cite{kim2016minimax}.
Estimating the 
instrinsic dimension when the data
are only approximately supported on a lower dimensional set
is much harder than the case
where ther support is precisely a lower dimensional set.

\subsection{Ridges}
\label{section::ridges}

Most manifold learning methods
assume that the distribution $P$ is supported
on some manifold $S$.
This is a very strong and unrealistic assumption.
A weaker assumption is that 
there may exist some low dimensional sets
where the density $p$ has a relatively high local concentration.
One way to make this more precise is through
the idea of density ridges.

A {\em density ridge} is a low dimensional set with large density.
But the distribution $P$ may not even have a density.
To deal with this issue, we define the smoothed distribution
$P_h$ obtained by convolving $P$ with a Gaussian.
Specifically, $P_h$ is the distribution with density
$$
p_h(x) = \int  K_h\left( x-u\right) dP(u)
$$
where $K_h(x) = h^{-d}(2\pi)^{-d/2}e^{-||x||^2/(2h^2)}$.
Note that $p_h$ is the mean of the kernel density estimator with bandwidth $h$.
The smoothed distribution $P_h$
always has a density, even if $P$ does not.
In topological inference,
we imagine using a small but positive $h$.
It is not necessary to let $h$ tend to 0 as we usual do in density estimation.
The salient topological features of $P$ will be preserved by $P_h$.

\begin{figure}
\begin{center}
\includegraphics[scale=.5]{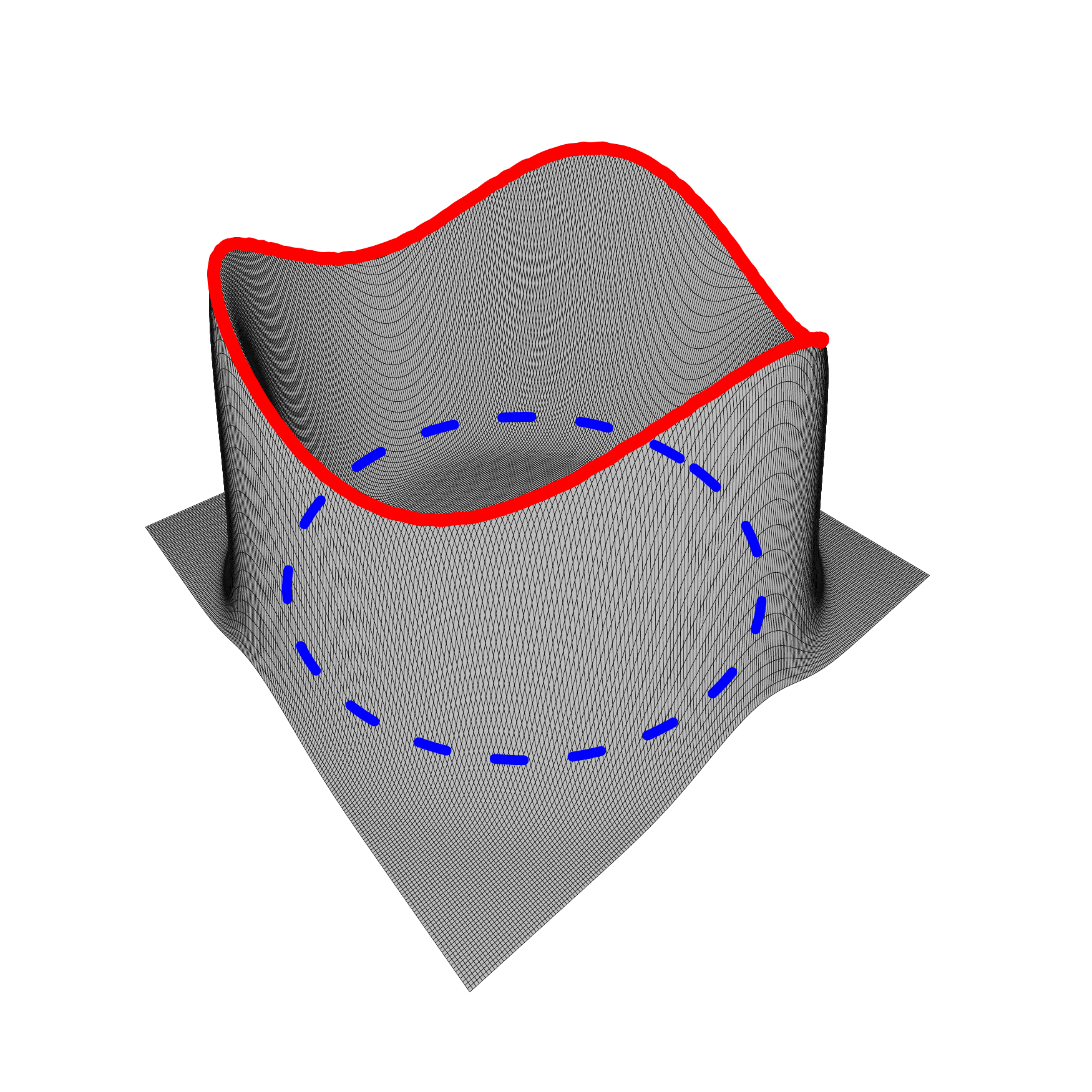}
\end{center}
\caption{This is a plot of a two-dimensional density function with a clearly defined
one-dimensional ridge. The ridge is the blue circle.}
\label{fig::circle}
\end{figure}

Let $g_h$ be the gradient of $p_h$ and let
$H_h$ be the Hessian.
Recall that a mode
of $p_h$ is a point $x$ with
$g_h(x)=(0,\ldots, 0)^T$ and
$\lambda_1(H_h(x)) < 0$.
A mode is a 0-dimensonal ridge.
More generally,
an $r$-dimensional ridge
is a set with sharp density in some directions, much like
the ridge of a mountain.
see Figure \ref{fig::circle}.
In fact, there are many ways to define a ridge;
see \cite{eberly1996ridges}.
We use the following defintion.
At a point $x$ we will define a local tangent space of
dimension $r$ and local normal space of dimension $d-r$.
Then $x$ is a ridge point if it is a local mode in the direction of the normal.
More precisely,
let
$\lambda_1(x)\geq \cdots \lambda_d(x)$
be the eigenvalues of the Hessian $H(x)$
and let
$v_1(x),\ldots, v_d(x)$ be the correspdonding
eigenvectors.
Let $V(x) = [v_{r+1}(x) \cdots v_d(x)]$ and define the projected gradient of $p$ by
$$
G(x) = V(x) V(x)^T g(x).
$$
The $r$-ridge is
$$
R_r(p) = \{x:\ G(x) =0,\ \lambda_{r+1}(x) <0 \}.
$$
Under suitable regularity conditions, this is indeed an $r$-dimensional set.

\begin{figure}
\vspace{-.5in}
\begin{center}
\includegraphics[scale=.4]{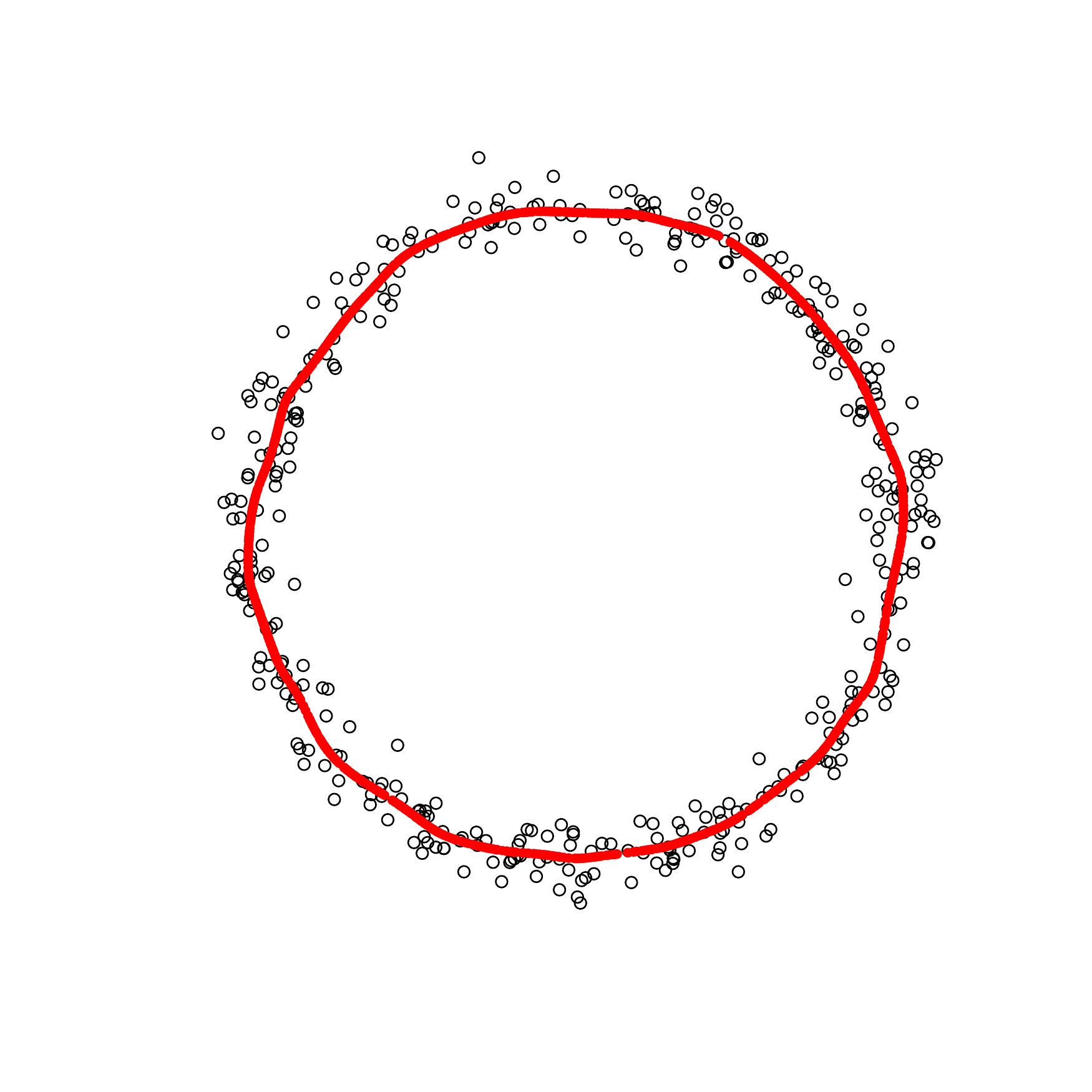}
\end{center}
\caption{The data are generated as $Y_i = X_i + \epsilon_i$
where the $X_i$ are sampled from a circle and $\epsilon_i$ are bivariate Gaussian.
The ridge $\hat R$ of the kernel density estimator is found using the SCMS algorithm
and is shown in red.}
\label{fig::scms}
\end{figure}

The ridge can be estimated by the ridge of a kernel density estimate.
Specifically, we take $\hat R = R_r(\hat p_h)$
to be the ridge of the kernel estimator.
The properties of this estimator are studied in
\cite{genovese2014nonparametric}
and
\cite{chen2015asymptotic}.
An algorithm for finding the ridge set of $\hat p_h$ was given by 
\cite{ozertem2011locally}
and is called the SCMS (subspace constrained mean shift algorithm).
Examples are shown in 
Figure \ref{fig::web} and
Figure \ref{fig::scms}.
A further example is in Section \ref{section::applications}.

Ridges can be related to manifolds as follows
(\cite{genovese2014nonparametric}).
Suppose we observe
$Y_1,\ldots, Y_n$ where
$Y_i = X_i + \sigma\epsilon_i$,
$X_1,\ldots, X_n \sim G$ is a sample from a distribution $G$ supported on 
a manifold $S$
and $\epsilon_1,\ldots, \epsilon_n$ are a sample from a noise distribution
such as a Gaussian.
As mentioned earlier,
$S$ can only
be estimated at a logarithmic rate.
However, if $\sigma$ is small enough and $S$ has positive reach,
then the density $p$ of $Y$ will have a well defined ridge $R$
such that
$H(R,S) = O(\sigma)$.
Furthermore, $R$ is ``topologically similar'' to
$S$ in a certain sense described in \cite{genovese2014nonparametric}.
In fact, $p_h$ will have a ridge $R_h$ such that
$H(R_h,S) = O(\sigma+h)$
and $R_h$ can be estimated at rate $O_P(\sqrt{\log n/n})$
independently of the dimension.

An example is shown in Figure \ref{fig::scms}.
The data are generated as follows.
We sample
$X_1,\dots, X_n$ uniform form a circle.
Then we set
$Y_i = X_i + \epsilon_i$
where $\epsilon_1,\ldots, \epsilon_n$
are draws from a bivariate Normal with mean $(0,0)$.
Next we find the kernel density estimator 
based on $Y_1,\ldots, Y_n$ and we find the ridge $\hat R$
of the kernel estimator using the SCMS algorithm.
The data are the black points in the plot.
The estimated ridge is shown in red.
Notice that the data are full dimensional but the estimated ridge
is one dimensional.

\subsection{Stratified Spaces}

Another generalization of manifold learning
is to assume that the support of $P$
is a {\em stratified space}
which means that the space can be decomposed into several,
intersecting submanifolds.
Estimation of stratified spaces is much less developed than manifold estimation.
Some examples include 
\cite{bendich2007inferring, skraba2014approximating}
and
\cite{bendich2007inferring}.
Ridge based methods
as discussed in Section \ref{section::ridges}
seem to work well
in this case
but, so far, this has not been established
theoretically.
A promising new approach
due to
\cite{arias2011spectral}
is based on a version of local PCA.

\section{PERSISTENT HOMOLOGY}
\label{section::persistence}

Persistent homology is a multiscale approach
to quantifying topological features in data
\citep{edelsbrunner2010computational,edelsbrunner2002topological,
edelsbrunner2008persistent}.
This is the branch of TDA that gets the most attention
and some researchers view TDA and
persistent homology as synonymous.

\begin{figure}
\begin{center}
\mbox{
\includegraphics[scale=.2]{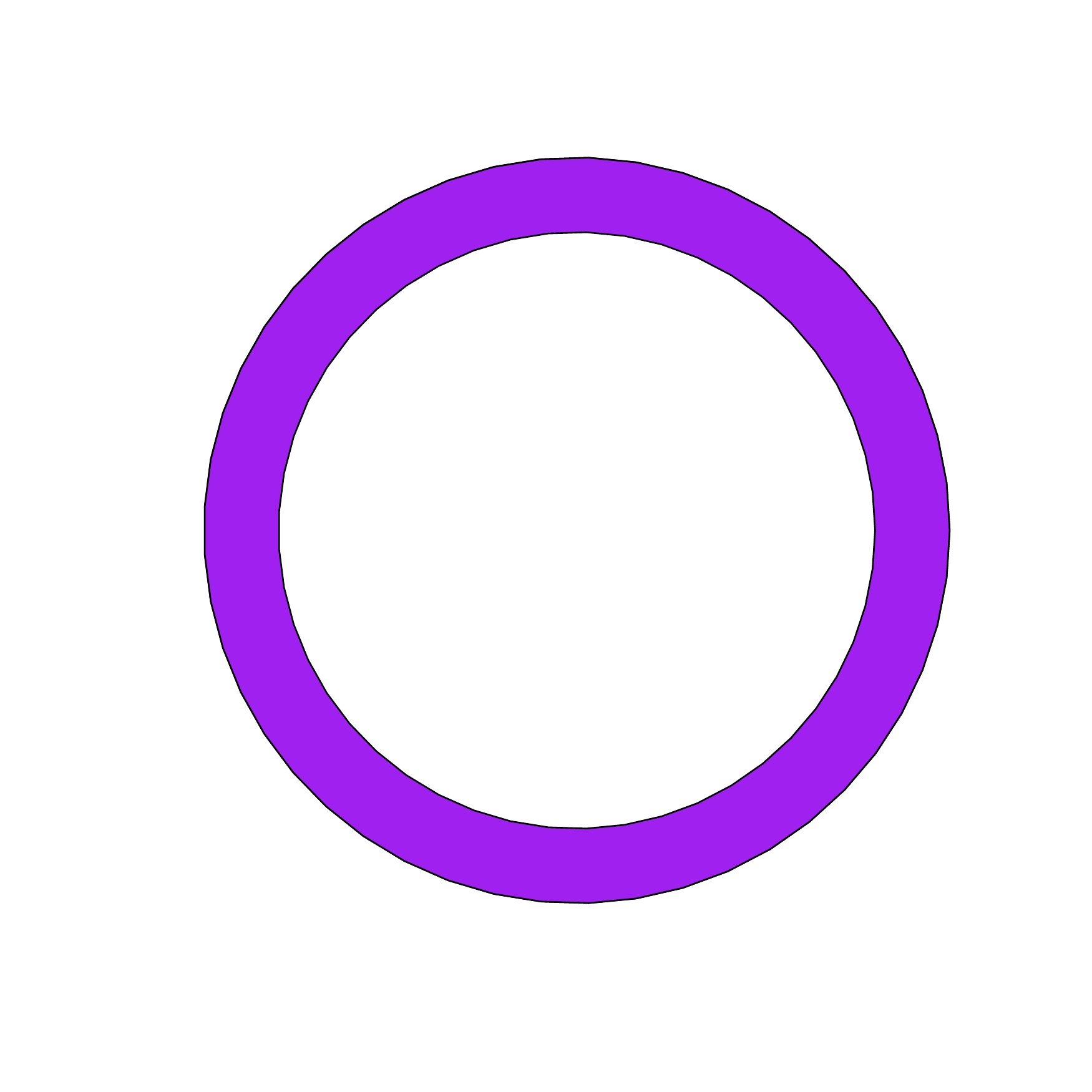}
\includegraphics[scale=.2]{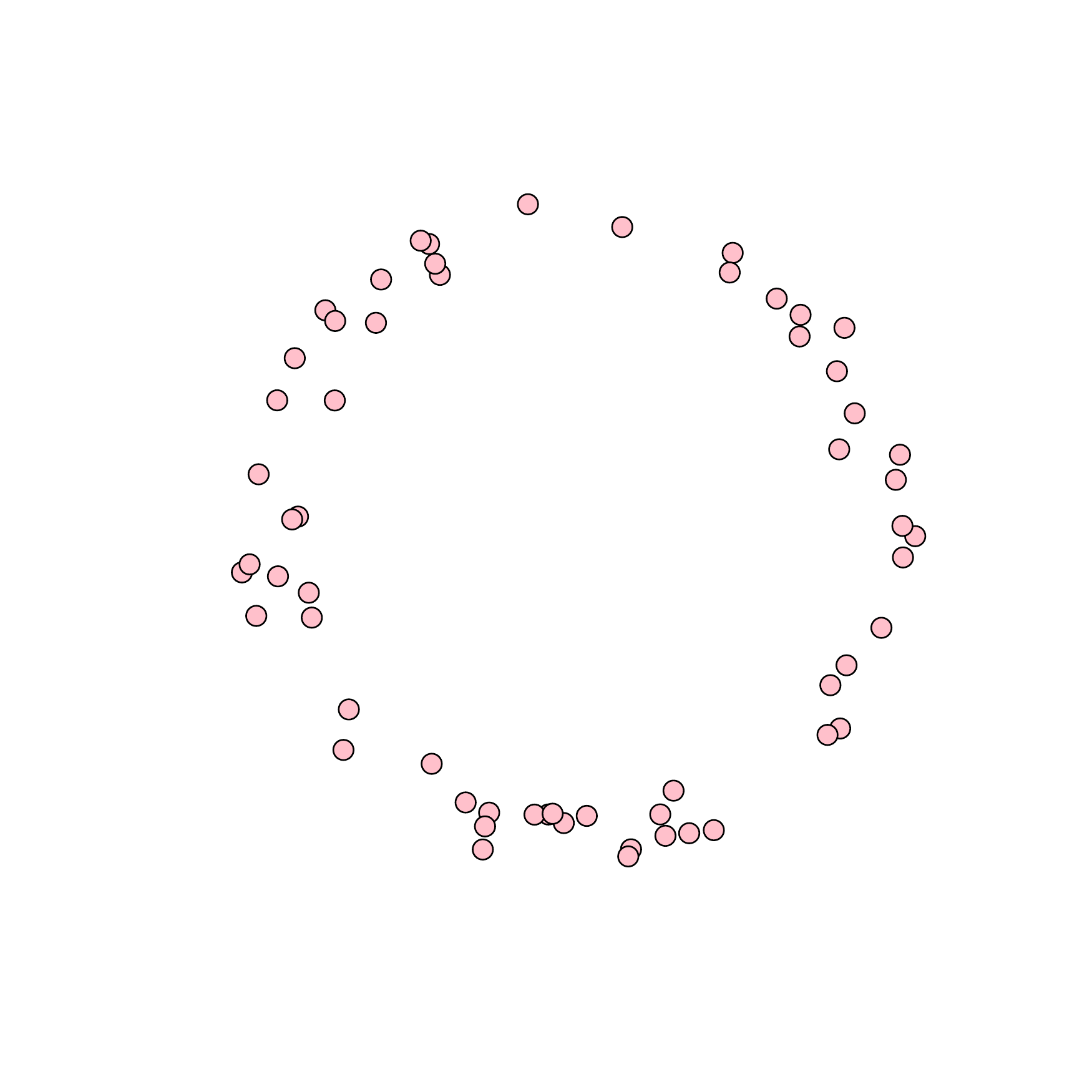}
\includegraphics[scale=.2]{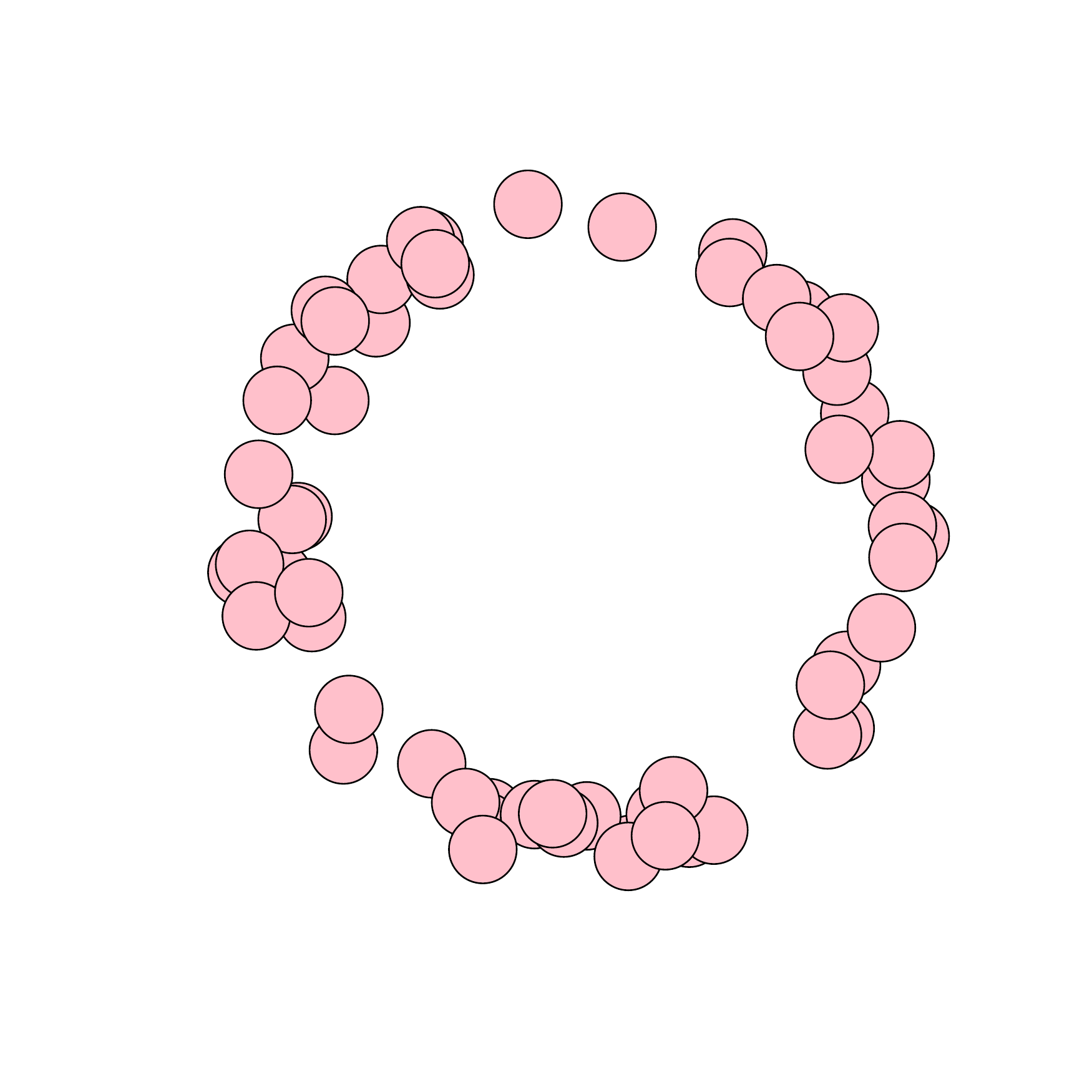}
\includegraphics[scale=.2]{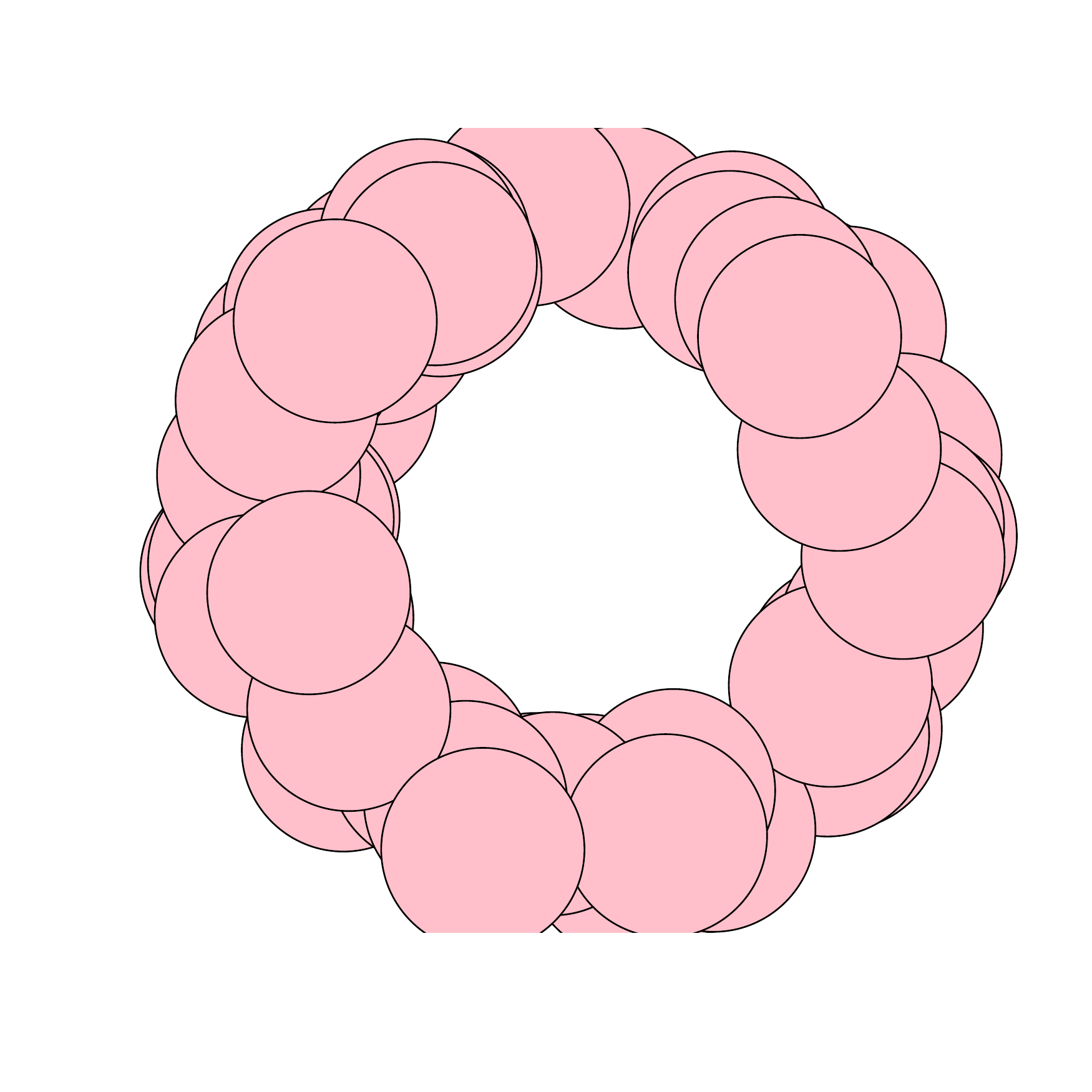}
}
\end{center}
\caption{Plot 1: the support $S$ of the distribution.
Plots 2-4: Union of balls $\bigcup_{i=1}^n B(X_i,\epsilon)$
around 60 data points drawn from a uniform on $S$,
with $\epsilon = 0.03,0.10, 0.30$.}
\label{fig::donut1}
\end{figure}

\begin{figure}
\begin{center}
\mbox{
\includegraphics[scale=.3]{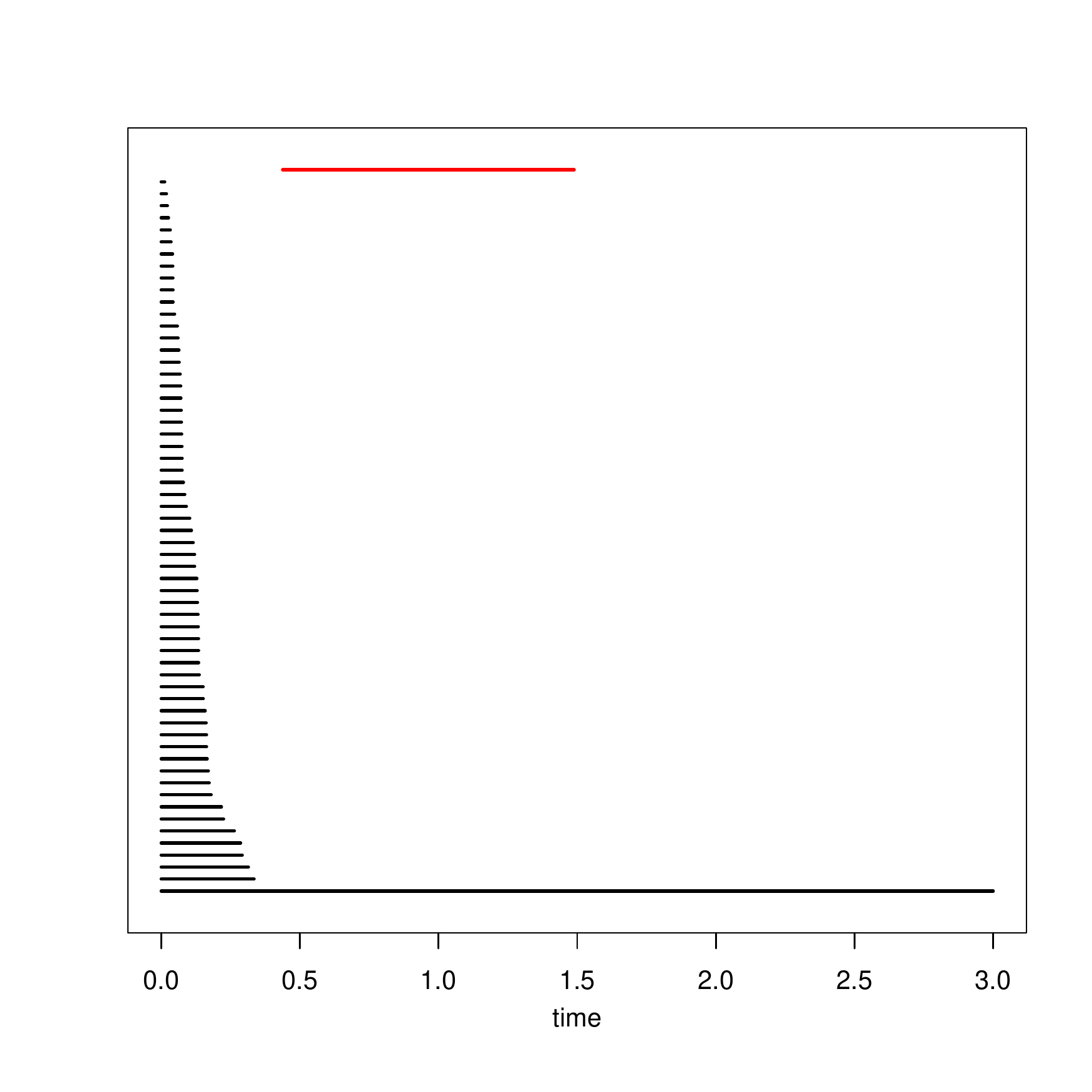}
\includegraphics[scale=.3]{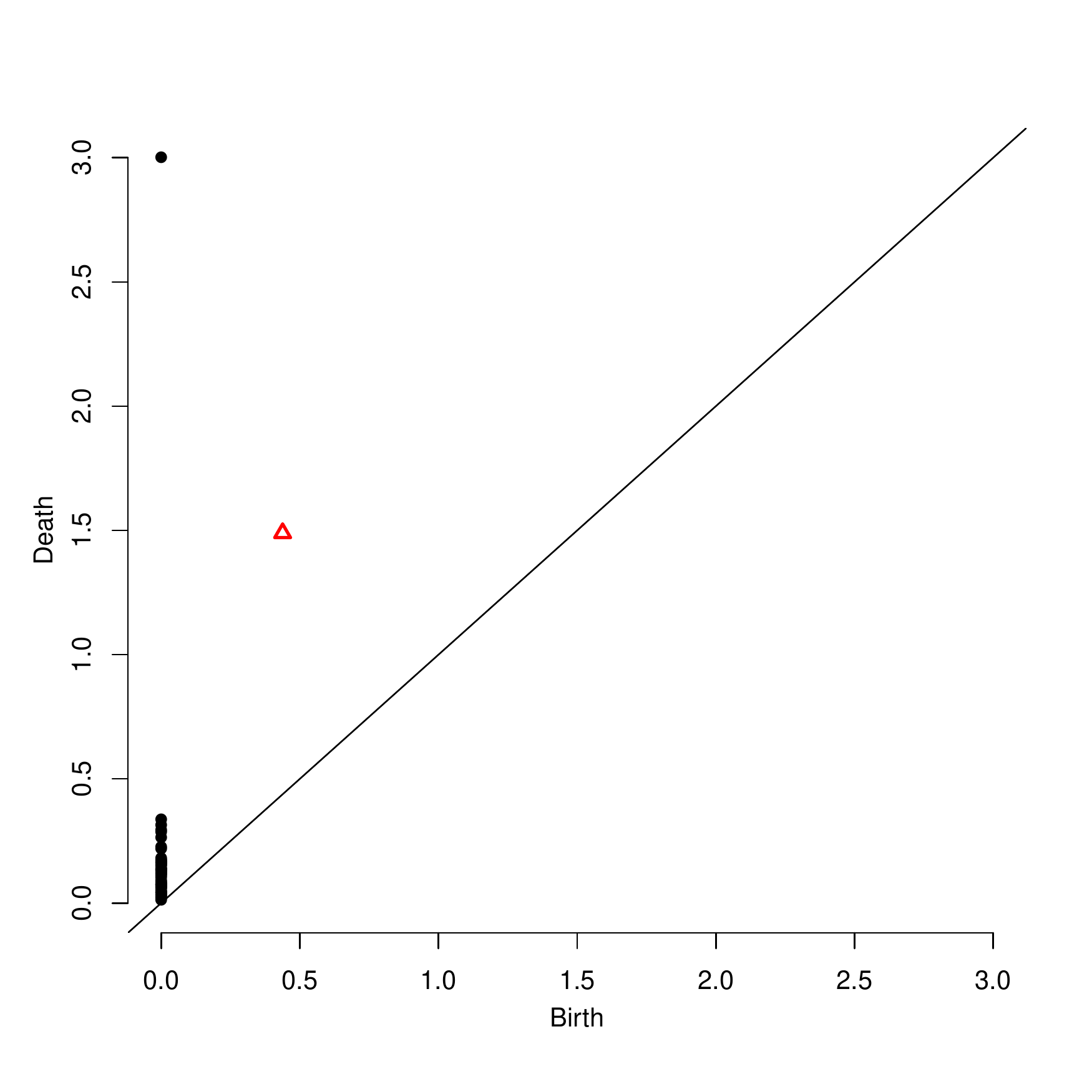}
}
\end{center}
\caption{Left: the barcode plot corresponding to the data from Figure \ref{fig::donut1}.
The black lines show the birth and death of each connected component as $\epsilon$ increases.
The red line shows the birth and death of the hole as $\epsilon$ increases.
Right: the persistence diagram.
In this case, the birth and death time of each feature is represented by a point on the diagram.
The black points correspond to connected components.
The red triangle corresponds to the hole.
Points close to the diagonal have a short lifetime.}
\label{fig::donut2}
\end{figure}

A quick, intuitive idea of persistent holomogy is given in Figures
\ref{fig::donut1} and \ref{fig::donut2}.
Here, we see some data and we also see the set
$\bigcup_{i=1}^n B(X_i,\epsilon)$
for various values of $\epsilon$.
The key observation is the topological features
appear and disappear as $\epsilon$ increases.
For example, when $\epsilon=0$ there are $n$ connected components.
As $\epsilon$ increases some of the connected components die
(that is, they merge) until only one connected component remains.
Similarly, at a certain value of $\epsilon$,
a hole is born.
The hole dies at a larger value of $\epsilon$.

Thus, each feature has a birth time and a death time.
The left plot in Figure \ref{fig::donut2}
is a {\em barcode plot}
which represents the birth time and death time of each feature
as a bar.
The right plot is a {\em persistence diagram}
where each feature is a point on the diagram
and the coordinates of the points are the birth time and death time.
Features with a long lifetime correspond to points far from the diagonal.
With this simple example in mind,
we delve into more detail.

\subsection{Homology}
\label{section::homology}

It is not possible to give a thorough review of homology
given the present space constraints.
But we can give a short, intuitive description
which will suffice for what follows.
More details are in the appendix and in 
\cite{fasy2014confidence}.
Good introductions can be found in
\cite{hatcher2000algebraic}
and \cite{edelsbrunner2010computational}.

\begin{figure}
\begin{center}
\mbox{
\begin{tikzpicture}[scale=.5]
\draw [fill=purple](5,0) circle (1in);
\draw [fill=white](5,0) circle (.5in);
\draw [fill=purple,purple](8.2,0) circle (.7in);
\draw [fill=white](8.2,0) circle (.2in);
\draw (.5,-3) -- (11.5,-3) -- (11.5,3) -- (.5,3) -- (.5,-3);
\end{tikzpicture}
\begin{tikzpicture}[scale=.5]
\draw [fill=purple](5,0) circle (1in);
\draw [fill=white](5,0) circle (.5in);
\draw [fill=purple,purple](10,0) circle (.7in);
\draw (1,-3) -- (12,-3) -- (12,3) -- (1,3) -- (1,-3);
\end{tikzpicture}
}
\end{center}
\caption{
The set on the left has one connected component and two holes and hence
$\beta_0 = 1$ and $\beta_1 = 2$.
The set on the right has two connected components and one hole
and hence $\beta_0 = 2$ and $\beta_1 = 1$.}
\label{fig::hom2}
\end{figure}
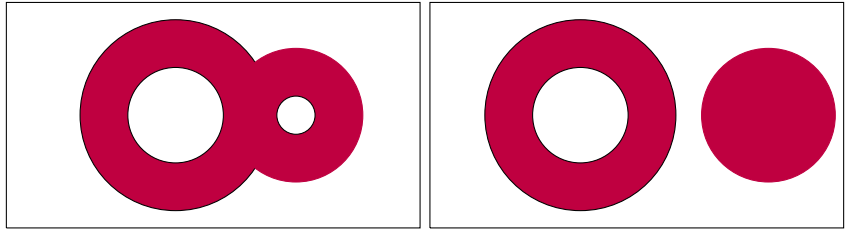

Homology characterizes sets based on
connected components and holes.
Consider the set on the left in Figure \ref{fig::hom2}.
The set has one connected component and two holes.
We write 
$\beta_0=1$ and $\beta_1=2$.
The numbers
$\beta_0,\beta_1,\ldots$
are called {\em Betti numbers}.
Intuitively,
$\beta_0$ is the number of connected components,
$\beta_1$ is the number of one-dimensional holes,
$\beta_2$ is the number of two-dimensional holes, etc.
(More formally, $\beta_j$ is the rank of the $j^{\rm th}$ homology group.)
The set on the right in Figure \ref{fig::hom2} has
two connected components and one hole, thus,
$\beta_0=2$ and $\beta_1=1$.
These holes are one-dimensional: they can be surrounded by a loop
(like a piece of string).
The inside of a soccer ball is a two dimensional hole.
To surround it, we need a surface.
For a soccer ball,
$\beta_0 = 1, \beta_1=0, \beta_2 =1$.
For a torus (a hollowed out donut),
$\beta_0 = 1, \beta_1=2, \beta_2 =1$.

The formal definition of homology
uses the language of group theory.
(The equivalence class of loops surrounding a hole
have a group structure.)
The details are not needed to understand the rest
of this paper.
Persistent homology examines these homological features from
a multiscale perspective.

\subsection{Distance Functions and Persistent Homology}

A good starting point for explaining persistent homology
is the {\em distance function}.
Given a set $S$, the distance function is defined to be
$$
d_S(x) = \inf_{y\in S}||x-y||.
$$
The lower level sets of the distance function are
$$
L_\epsilon = \{ x:\ d_S(x) \leq \epsilon\}.
$$
We also have that
$$
L_\epsilon = \bigcup_{x\in S}B(x,\epsilon).
$$
So $L_\epsilon$ can be thought of either as a union of balls,
or as the lower level set of the distance function.
As $\epsilon$ increases, the sets $L_\epsilon$ evolve.
Topological features --- connected components and holes ---
will appear and disappear.
Consider the circle
$$
S = \{ (x,y):\ x^2 + y^2 = 1\}.
$$
The set $L_\epsilon$ is an annulus of radius $\epsilon$.
For all values of $\epsilon$, $L_\epsilon$ has one connected component.
For $0\leq \epsilon < 1$, the set $L_\epsilon$ has one hole.
The hole dies at $\epsilon=1$.
Thus, the hole has birthtime $\epsilon=0$ and deathtime $\epsilon=1$.
In general, these features can be represented as a persistence diagram $D$ as in
Figure \ref{fig::donut2}.
The diagram $D$ represents the persistent homology of $S$.

\begin{figure}
\begin{center}
\includegraphics[scale=.5]{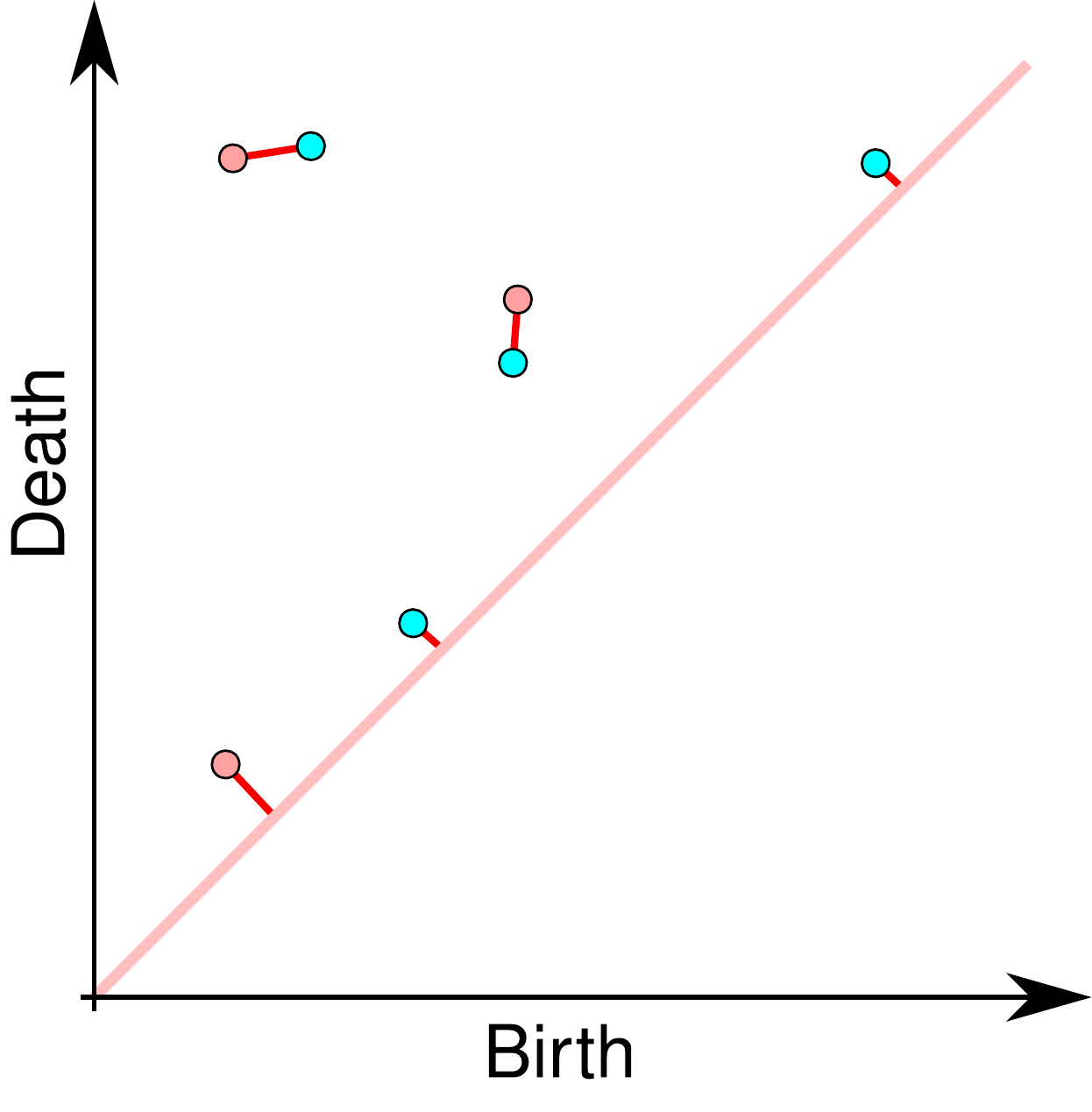}
\end{center}
\caption{The bottleneck distance between two persistence diagrams is computed
by finding the best matching between the two diagrams.
This plot shows two diagrams that have been overlayed.
The matching is indicated by the lines joining the points from the two diagrams.
Note that some points --- those with short lifetimes --- are matched to the diagonal.}
\label{fig::bottleneck}
\end{figure}

Technically, the persistence diagram $D$ 
is a multiset consisting of all pairs of points on the plot
as well as all points on the diagonal.
Given two diagrams $D_1$ and $D_2$,
the {\em bottleneck distance} defined by
\begin{equation}
\delta_\infty(D_1,D_2) = \inf_\gamma \sup_{z\in D_1}||z - \gamma(z)||_\infty
\end{equation}
where $\gamma$ ranges over all bijections between $D_1$ and $D_2$.
Intuitively, this is like overlaying the 
two diagrams and asking how much we have to shift the points on the diagrams to make them the same.
See Figure \ref{fig::bottleneck}.

Now suppose we observe a sample
$X_1,\ldots, X_n$ drawn from a distribution $P$
supported on $S$.
The {\em empirical distance function} is
$$
\hat d(x) = \min_{1\leq i\leq n} ||x-X_i||.
$$
Note that the lower level sets of $\hat d$ are precisely the union of balls
described in the last section:
$$
\hat L_\epsilon = \{x:\ \hat d(x) \leq \epsilon\} = \bigcup_{i=1}^n B(X_i,\epsilon).
$$
The persistence diagram $\hat D$ defined by these lower level sets
is an estimate of the underlying diagram $D$.

The empirical distance function is the most commonly used
method for defining the persistence diagram of a dataset
in the field of computational topology.
But from a statistical point of view,
this is a very poor choice.
It is clear that $\hat d$ is highly non-robust.
Even a few outliers will play havoc with the estimator.

Fortunately,
more robust and statistically sound methods
are available.
The first, and perhaps most natural for statisticians,
is to replace the lower level sets of the empirical distance function,
with the upper level sets of a density estimator.
This approach has been suggested by
\cite{phillips2015geometric,
chazal2014robust,
bobrowski2014topological,
chung2009persistence, bubenik2015statistical}.
The idea is to
consider
the upper level sets
$\hat L_t = \{x:\ \hat p_h(x) > t\}$.
As $t$ varies from $\sup_x \hat p_h(x)$ down to 0,
the sets $\hat L_t$
evolve and the birth and death times
of features are again recorded on a persistence diagram.
In this case, 
the birth times are actually after the death times.
This is just an artifact from using upper level sets
instead of lower level sets.

An alternative is to re-define the distance function to be
intrinsically more robust.
Specifically, \cite{chazal2011geometric}
defined the 
{\em distance to a measure (DTM)} as follows.
Let $0 \leq m \leq 1$ be a scale parameter
and define
$$
d^2_m(x) = \frac{1}{m}\int_0^m \delta^2_a(x) da
$$
where
$$
\delta_a(x) = \inf\{r>0:\ P(B(x,r))> a\}.
$$
We can think of $d_m$ as a function $T(P)$ of the distribution $P$.
The plug-in estimate of $d_m$ obtained by inserting the empirical distribution in place of $P$ is
$$
\hat d_m^2 (x) = \frac{1}{k}\sum_{i=1}^k ||x - X_i(x)||^2
$$
where
$k = \lfloor mn\rfloor$
and
$X_j(x)$ denote the data after re-ordering them so that
$||X_1(x)-x|| \geq ||X_2(x) -m|| \geq \cdots$.
In other words,
$\hat d_m^2(x)$ is just the average squared distance to the
$k$-nearest neighbors.

The definition
of $d_m$ is not arbitrary.
The function $d_m$ preserves certain crucial properties
that the distance function has,
but it changes gracefully as we allow more and more noise.
It is essentially a smooth, probabilistic version of the distance function.
The properties of the DTM are discussed in
\cite{chazal2011geometric, chazal2014robust, chazal2015rates}.

Whether we use the kernel density estimator or the DTM,
we would like to have a way to decide when topological features are statistically significant.
\cite{fasy2014confidence,chazal2014robust} suggest the following method.
Let
$$
F(t) = P( \sqrt{n} \, \delta_\infty(\hat D,D)\leq t)
$$
where $D$ is the true diagram and $\hat D$ is the estimated diagram.
Any point on the diagram that is farther than
$t_\alpha=F^{-1}(1-\alpha)$ from the diagonal is considered significant
at level $\alpha$.
Of course, $F$ is not known but can be estimated by the bootstrap:
$$
\hat F(t) = \frac{1}{B}\sum_{j=1}^B I( \sqrt{n} d_\infty(\hat D^*_j,\hat D)\leq t)
$$
where
$\hat D^*_1, \ldots, \hat D^*_B$
are the diagrams based on $B$ bootstrap samples.
Then
$\hat t_\alpha= \hat F^{-1}(1-\alpha)$ is an estimate of $t_\alpha$.

\vspace{1cm}

\begin{figure}
\begin{center}
\includegraphics[scale=.3]{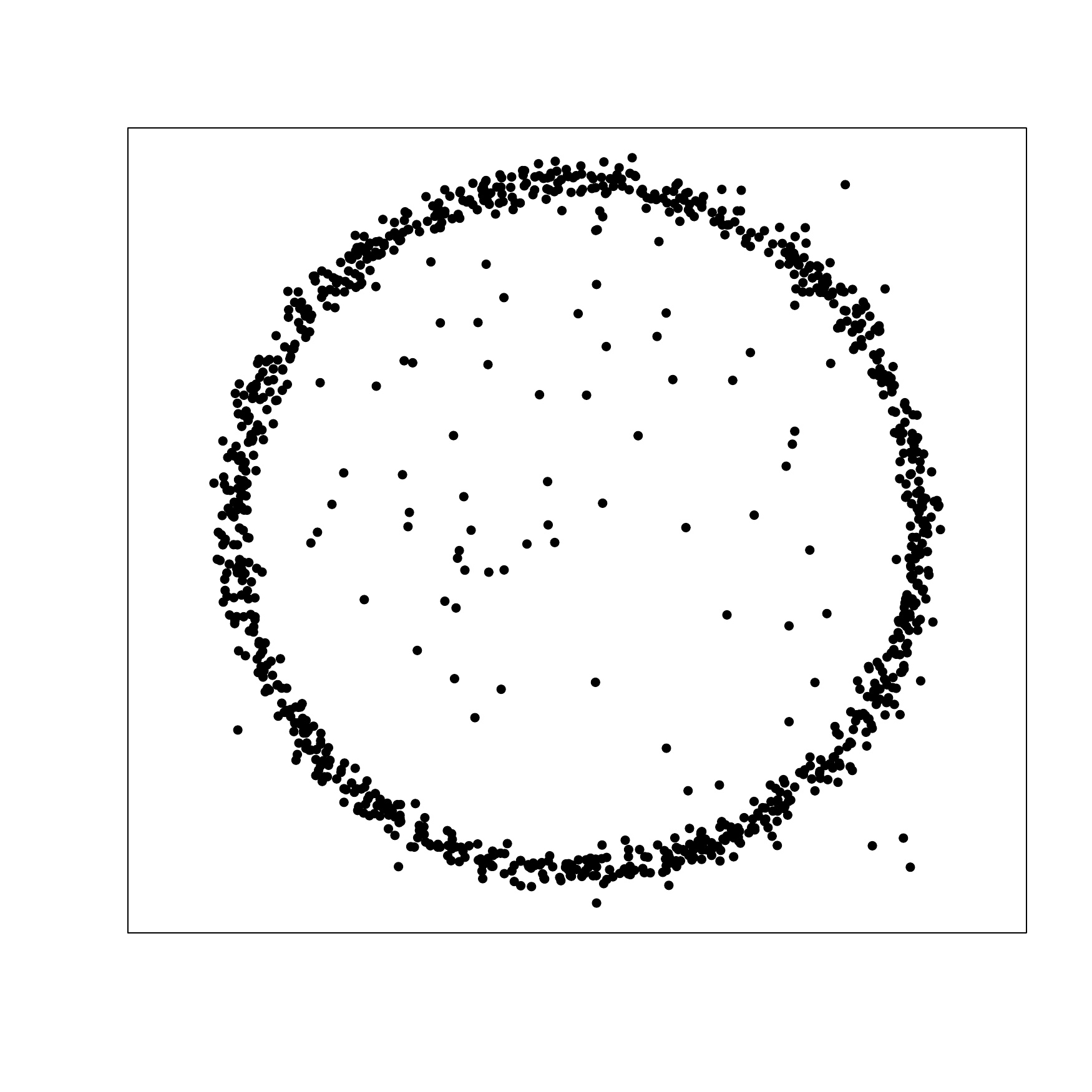}
\end{center}
\caption{Data sampled from a circle, with Gaussian noise added.
There are also 100 outliers sampled uniformly from the square.}
\label{fig::Density-and=DTM1}
\end{figure}

\begin{figure}
\begin{center}
\mbox{\includegraphics[scale=.3]{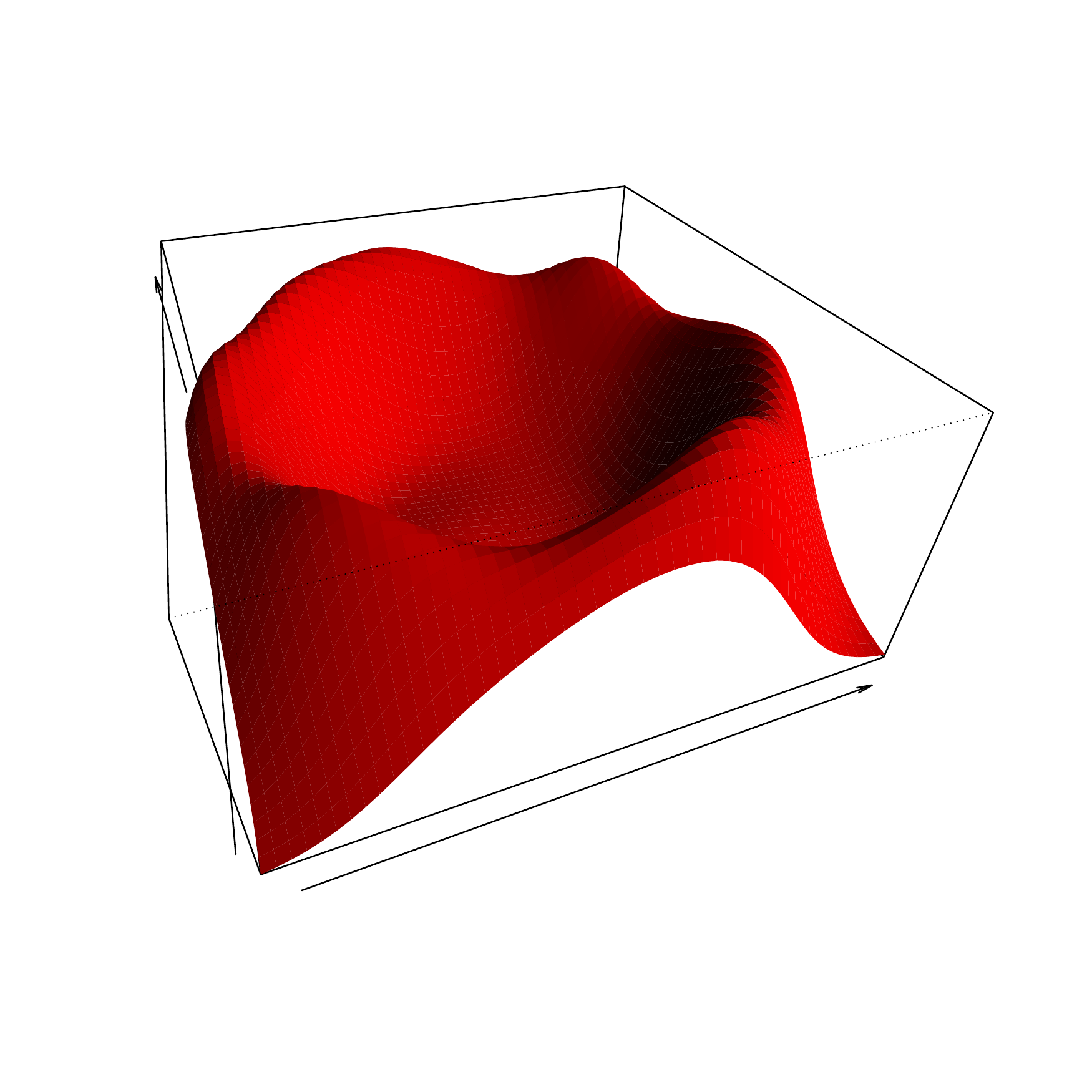}
\includegraphics[scale=.3]{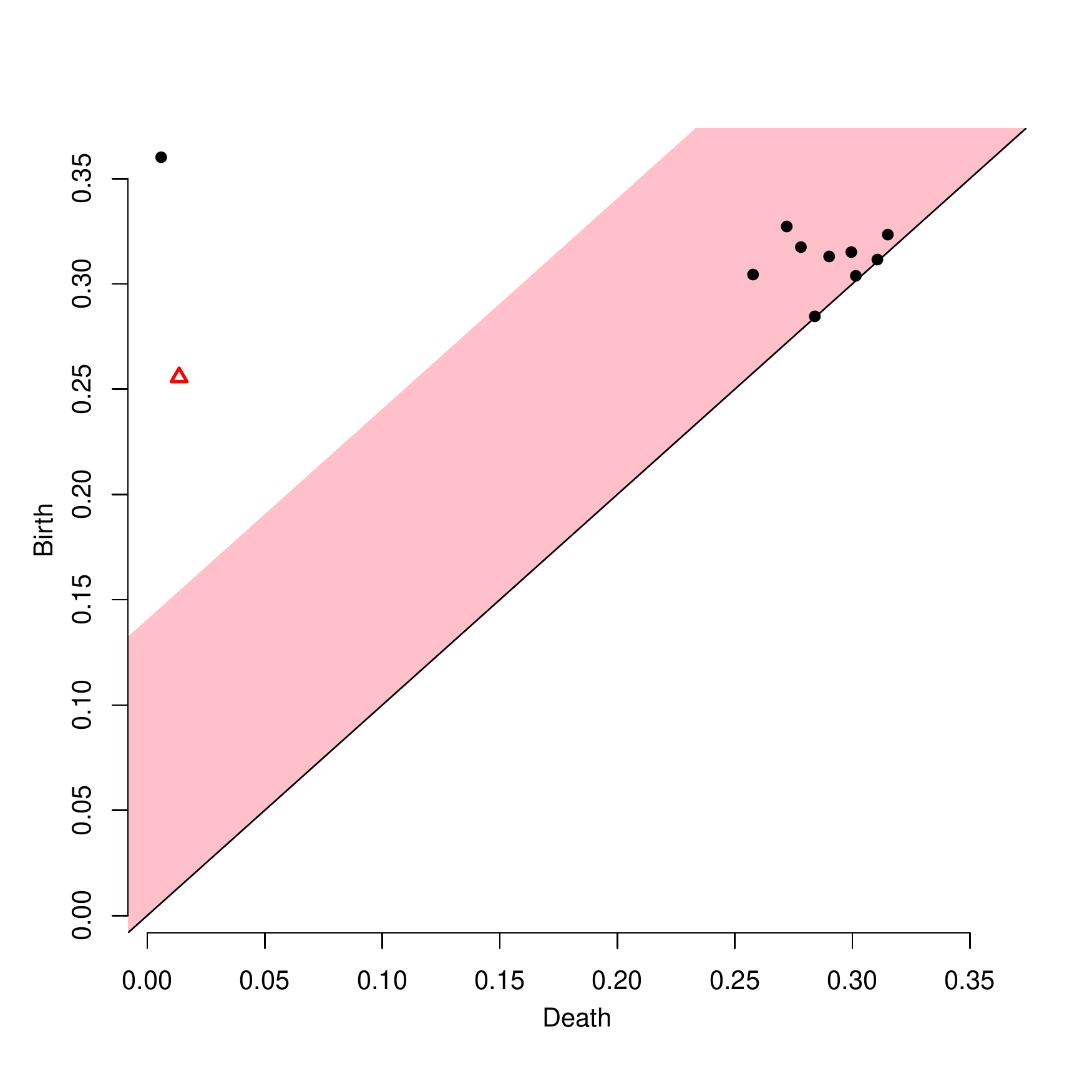}}
\end{center}
\caption{Left: the kernel density estimator.
Right: the persistence diagram corresponding to the upper level sets of the estimator.
The points above the pink band are significant compared to the bootstrap critical value.
Note that one connected component (the black dot)
and one hole (the red triangle) are significant.}
\label{fig::Density-and=DTM23}
\end{figure}

\begin{figure}
\begin{center}
\mbox{\includegraphics[scale=.3]{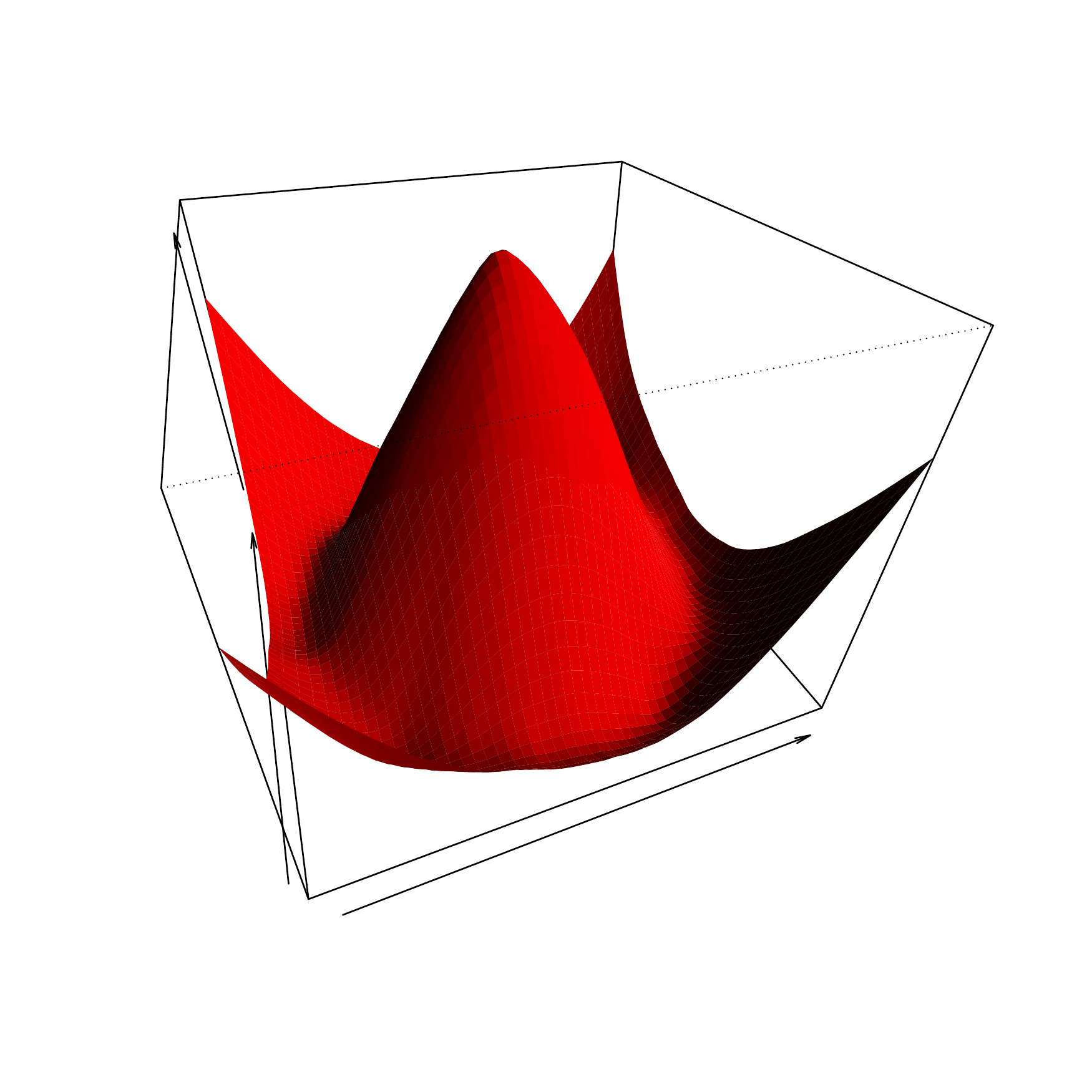}
\includegraphics[scale=.3]{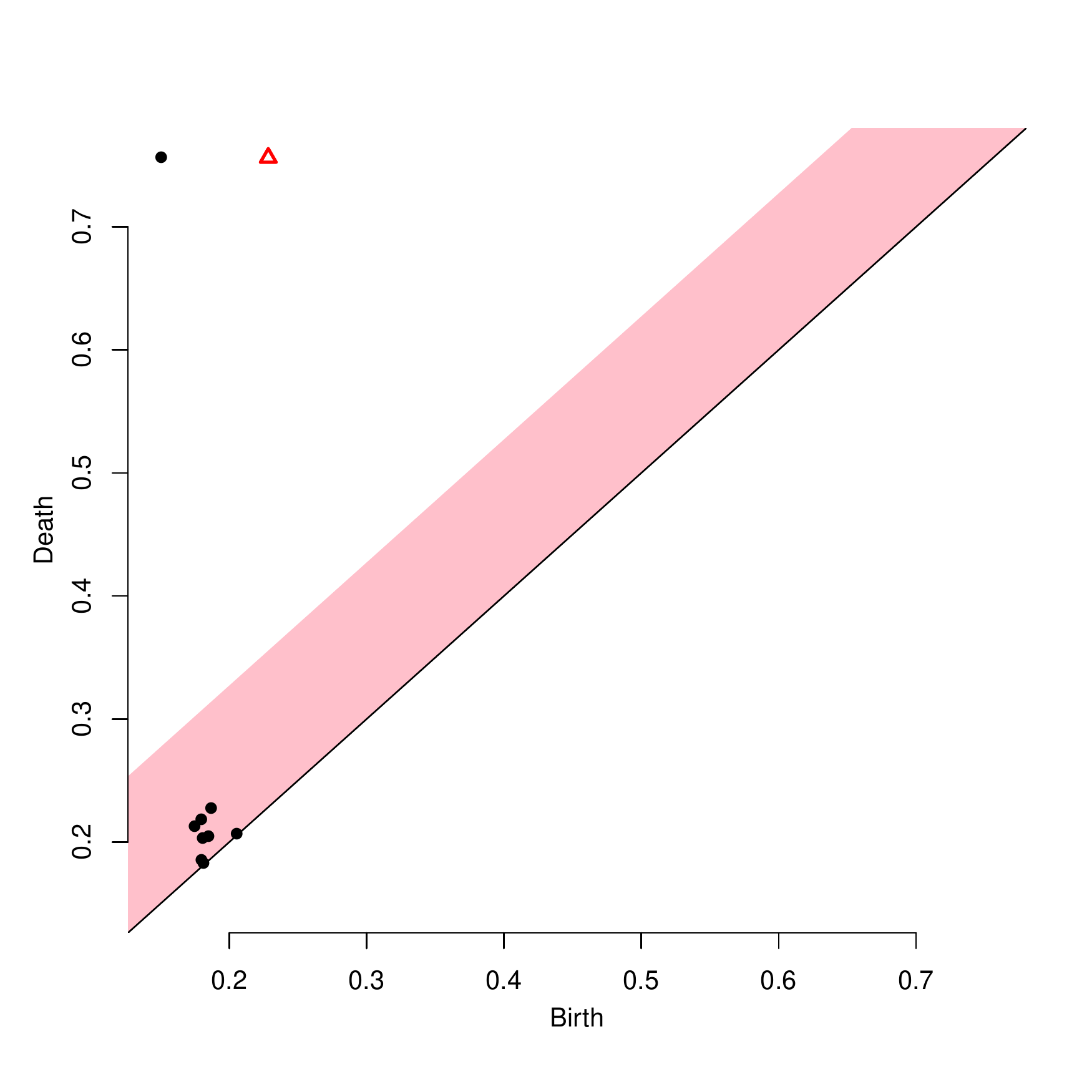}}
\end{center}
\caption{Left: the DTM.
Right: the persistence diagram corresponding to the lower level sets of the DTM.
The points above the pink band are significant compared to the bootstrap critical value.
Note that one connected component (the black dot)
and one hole (the red triangle) are significant.}
\label{fig::Density-and=DTM45}
\end{figure}

{\bf Example.}
We sampled 1,000 observations from a circle in $\mathbb{R}^2$.
Gaussian noise was then added to each observation.
Then we added 100 outliers samples uniformly from the square.
The data are shown in 
Figure \ref{fig::Density-and=DTM1}.
Figure \ref{fig::Density-and=DTM23} shows the kernel density estimator ($h=.02$)
and the persistence diagram based on the upper level sets of the estimator.
The points in the pink band are not significant at level $\alpha = 0.1$
(based on the bootstrap).
The two points that are significant correspond to one connected component (black dot) and one hole
(red triangle).
Figure \ref{fig::Density-and=DTM45} shows a similar analysis of the same data using the DTM with $m=.1$.
Generally, we find that the significant features
are more prominent using the DTM rather than the kernel density estimator.
Also, the DTM is less sensitive to the choice of tuning parameter
although it is not known why this is true.

\subsection{Simplicial Complexes}

The persistence diagram is not computed directly from
$\hat L_\epsilon$.
Instead, one forms an object called a
{\em \v{C}ech complex}.
The \v{C}ech complex $C_\epsilon$ is defined as follows.
All singletons are included in $C_\epsilon$; these are 0-dimensional simplices.
All pairs of points $X_i,X_j$ such that
$||X_i - X_j|| \leq \epsilon$ are included in $C_\epsilon$;
these are 1-dimensional simplices.
Each triple $X_i,X_j,X_k$ such that
$B(X_i,\epsilon/2)\cap B(X_j,\epsilon/2)\cap B(X_k,\epsilon/2)$ is non-empty,
is included in $C_\epsilon$; these are 2-dimensional simplices.
And so on.
The \v{C}ech complex is an example of a {\em simplicial complex}.
A collection of simplices is a simplicial complex if it
satisfies the following condition:
if $F$ is a simplex in $C_\epsilon$ and $E$ is a face of $F$, then $E$ is also on $C_\epsilon$.
It can be shown that the homology of
$\hat L_\epsilon$ is the same as the homology of $C_\epsilon$.
But the homology of $C_\epsilon$ can be computed using basic matrix
operations.
This is how homology is computed in practice
\citep{edelsbrunner2010computational}.
Persistent homology relates the complexes as $\epsilon$ varies.
Again, all the relevant computations can be reduced to linear algebra.
Working directly with the \v{C}ech complex is computationally prohibitive.
In practice, one often uses the
{\rm Vietoris-Rips complex} $V_\epsilon$ which is defined as follows.
A simplex is included in $V_\epsilon$ if each pair of vertices is no more than $\epsilon$ apart.
It can be shown that the persistent homology defined by $V_\epsilon$ approximates
the persistent homology defined by $C_\epsilon$.

\subsection{Back To Density Clustering}

\cite{chazal2013persistence}
have shown that persistent homology
can be used as a tool for density clustering.
This idea was futher examined in
\cite{genovese2016non}.
Thus we have come full circle and returned to 
the topic of Section \ref{sec::density}.

Recall the mode clustering method described
in Section \ref{sec::modes}.
We estimate the density, find the modes
$\hat m_1,\ldots, \hat m_k$
and the basins of attraction
$C_1,\ldots, C_k$
corresponding to the modes.

\begin{figure}
\begin{center}
\begin{tabular}{cc}
\includegraphics[scale=.3]{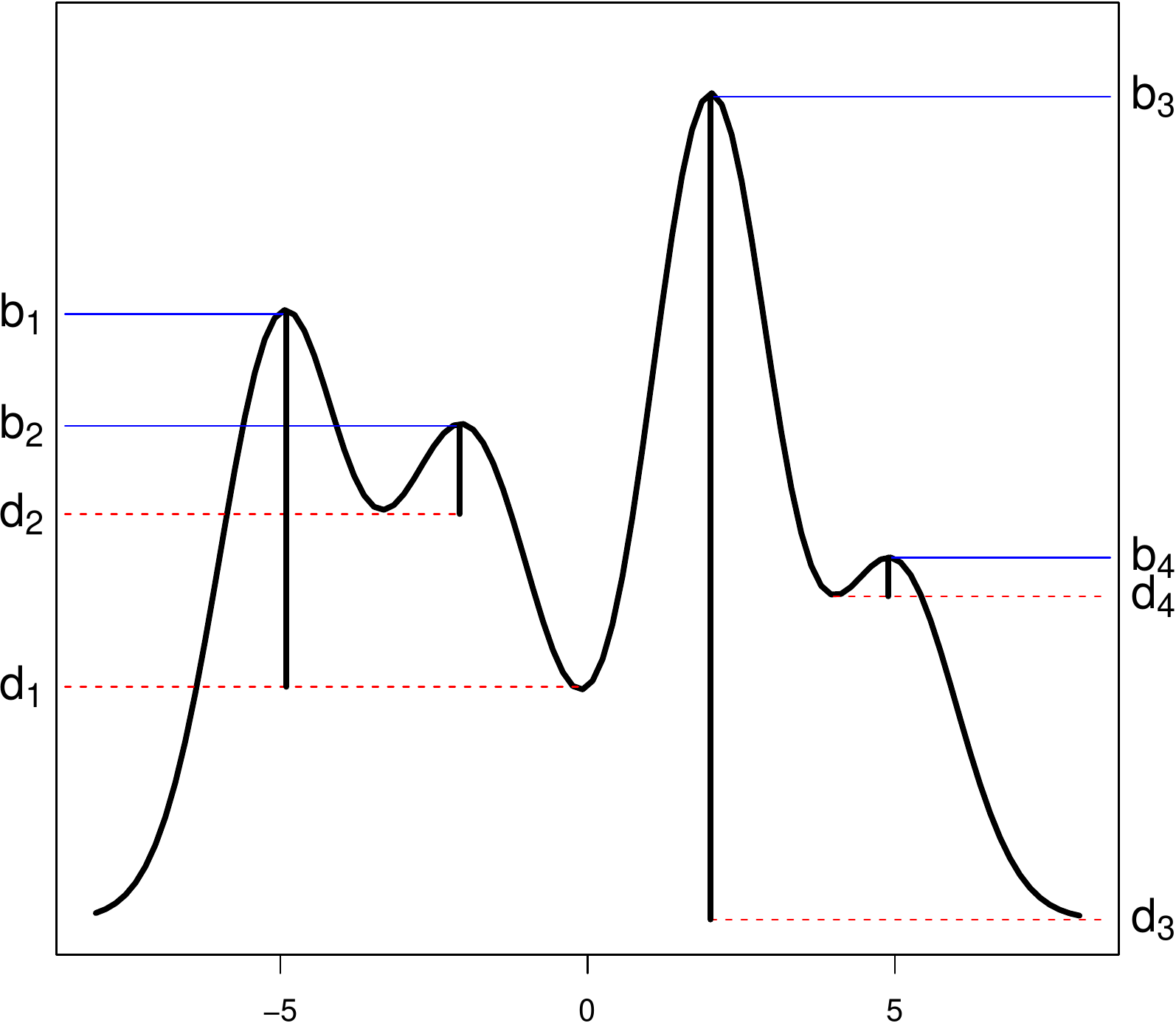}&
\includegraphics[scale=.3]{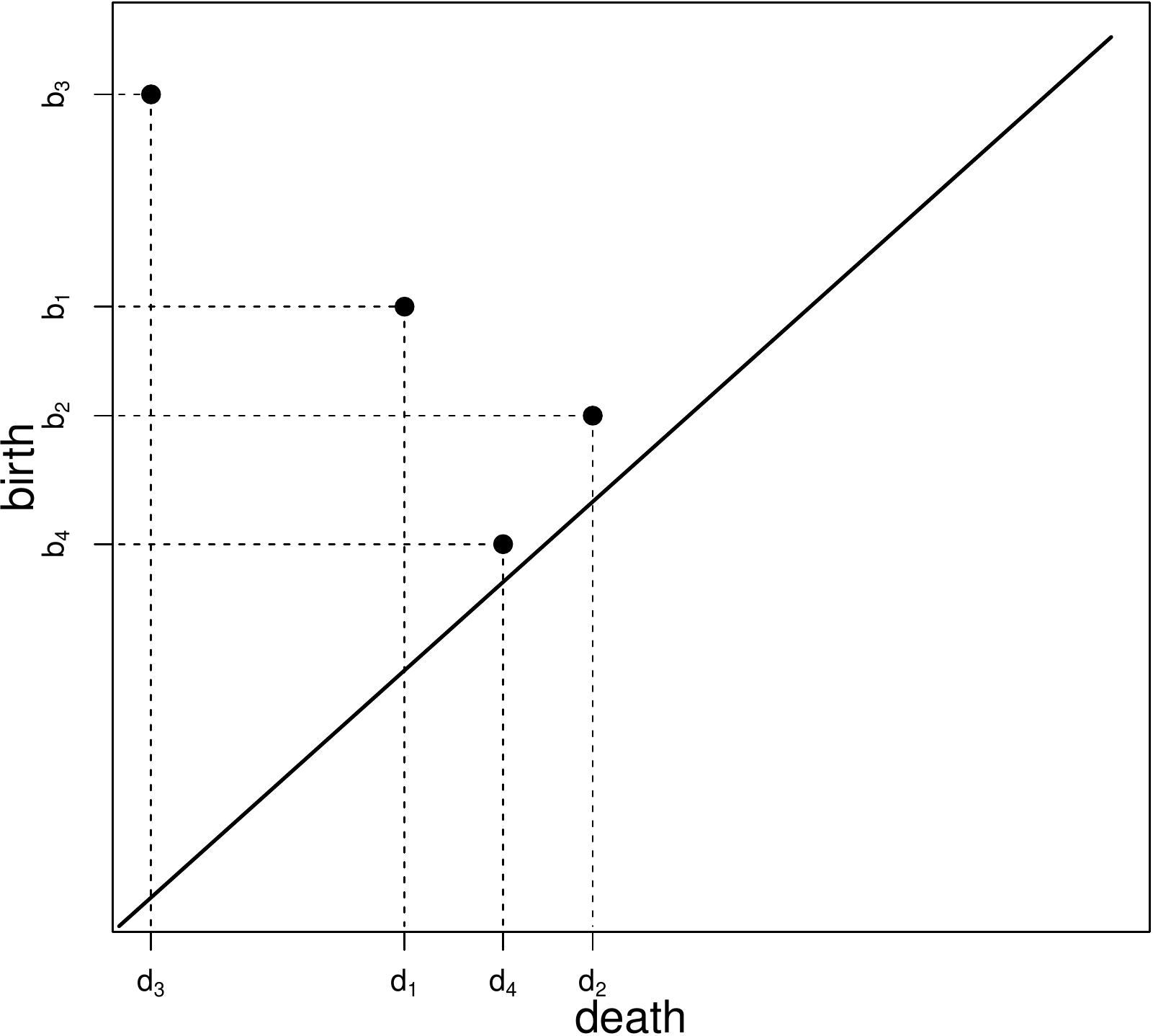}
\end{tabular}
\end{center}
\caption{\em Starting at the top of the density and moving down,
each mode has a birth time $b$ and a death time $d$.
The persistence diagram (right) plots the points
$(d_1,b_1),\ldots, (d_4,b_4)$.
Modes with a long lifetime are far from the diagonal.}
\label{fig::explain}
\end{figure}

But we can use more information.
In the language of persistent homology,
each mode has a lifetime.
See Figure \ref{fig::explain}.
Suppose we start with $t = \sup_x p(x)$.
We find the upper level set $L_t = \{x:\ p(x) \geq t\}$.
Now we let $t$ decrease.
(We can think of $t$ as ``time'' but, in this case, time 
runs backwards since it starts at a large number and tends to 0.)
Everytime we get to a new mode, a new connected component of $L_t$ is born.
However, as $t$ decreases, the connected components can merge.
When they merge, the most recently created component is considered to be dead
while the other component is still alive.
This is called the ``elder rule.''
Proceeding this way, small modes correspond to level sets with short lifetimes.
Strong modes correspond to level sets with long lifetimes.
We can plot the information as a persistence diagram as in the right plot of
Figure \ref{fig::explain}.

We can use this representation of the modes
to decide which modes of a density estimator are significant
\citep{chazal2014robust, chazal2013persistence}.
Define $\hat t_\alpha$ by
$$
\mathbb{P}\bigl(\sqrt{n}||\hat p_h^* - \hat p_h|| > \hat t_\alpha \bigm| X_1,\ldots, X_n\bigr) = \alpha,
$$
where
$\hat p_h^*$ is based on a bootstrap sample $X_1^*,\ldots, X_n^*$
drawn from the empirical distribution $P_n$.
The above probability can be estimated by
$$
\frac{1}{B}\sum_{j=1}^B I(\sqrt{n}||\hat p_h^* - \hat p_h|| > t)
$$
Any mode whose corresponding point on the persistence diagram
is farther than $\hat t_\alpha$ from the diagonal
is considered a significant mode.

\section{TUNING PARAMETERS AND LOSS FUNCTIONS}

Virtually every method we have discussed in this paper
requires the choice of a tuning parameter.
For example, many of the methods involve a kernel density estimator
which requires a bandwidth $h$.
But the usual methods for choosing tuning parameters
may not be appropriate for TDA.
In fact, the problem of choosing tuning parameters is one of
the biggest open challenges in TDA.

Let us consider the problem of
estimating a density $p$ with the kernel estimator $\hat p_h$.
The usual $L_2$ risk is
$\mathbb{E}[\int (\hat p_h(x) -p(x))^2 dx]$.
Under standard smoothness assumptions,
the optimal bandwidth
$h\asymp n^{-1/(4+d)}$ yielding a risk of order 
$n^{-4/(4+d)}$.

But in TDA we are interested in shape, not $L_2$ loss
(or $L_p$ loss for any $p$).
And, as I have mentioned earlier,
it may not even be necessary to let $h$ tend to 0
to capture the relevant shape information.
In Section \ref{section::trees}
we saw that, in some cases,
the density tree $T(p_h)$
has the same shape as the true tree $T(p)$
even for fixed $h>0$.
Here, $p_h(x)=\mathbb{E}[\hat p_h(x)]$.

Similarly, consider estimating a ridge $R$
of a density $p$.
In general, the ridge can only be estimated at rate
$O_P(n^{-2/(8+d)})$.
Now suppose we use a small but fixed (non-decreasing) bandwidth $h$.
Usually, the ridge $R_h$ of $p_h$ is a reasonably good but slightly biased approximation
to $R$. But $R$ can be estimated at rate
$O_P(\sqrt{\log n/n})$.
We are often better off living with the bias and estimating $R_h$ instead of $R$.

In fact one could argue that any shape information
that can only be recovered with small bandwidths
is very subtle and cannot be reliably estimated.
The salient structure can be recovered with a fixed bandwidth.
To explain this in more detail,
we consider two examples from
\cite{chen2015density}.

\begin{figure}
\begin{center}
\begin{tabular}{cc}
\includegraphics[scale=.3]{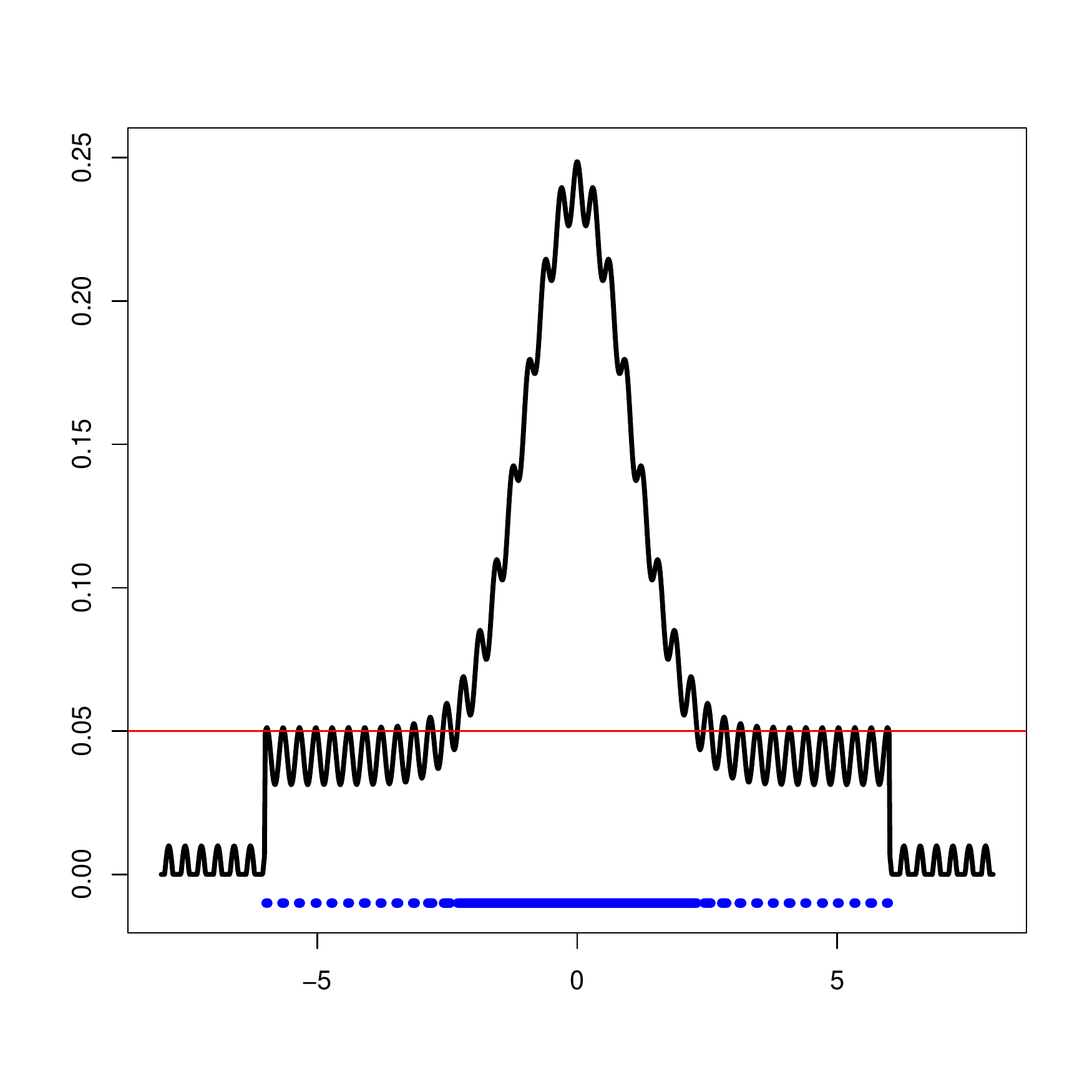} & 
\includegraphics[scale=.3]{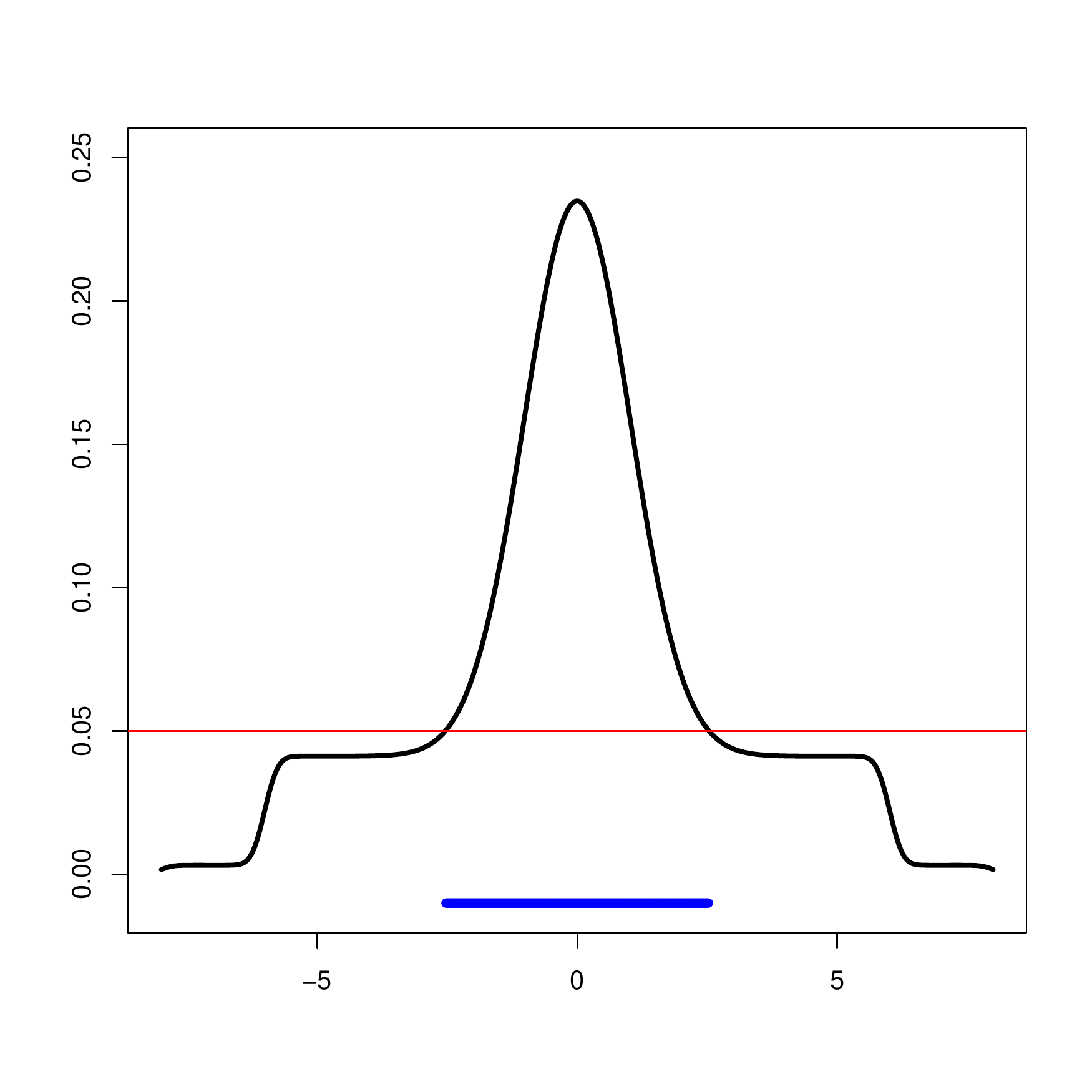}
\end{tabular}
\end{center}
\caption{Left: a density $p$ and a level set $\{p > t\}$.
Right: the smoothed density $p_h$ and the level set $\{p_h > t\}$.}
\label{fig::j1}
\end{figure}

The left plot in Figure
\ref{fig::j1} shows a density $p$.
The blue points at the bottom show the level set $L=\{x:\ p > .05\}$.
The right plot shows
$p_h$ for $h=.2$ and
the blue 
points at the bottom show the level set $L_h=\{x:\ p_h > .05\}$.
The smoothed out density $p_h$ is biased
and the level set $L_h$ loses the small details of $L$.
But $L_h$ contains the main part of $L$
and it may be more honest to say that
$\hat L_h$ is an estimate of $L_h$.

\begin{figure}
\begin{center}
\begin{tabular}{cc}
\includegraphics[scale=.3]{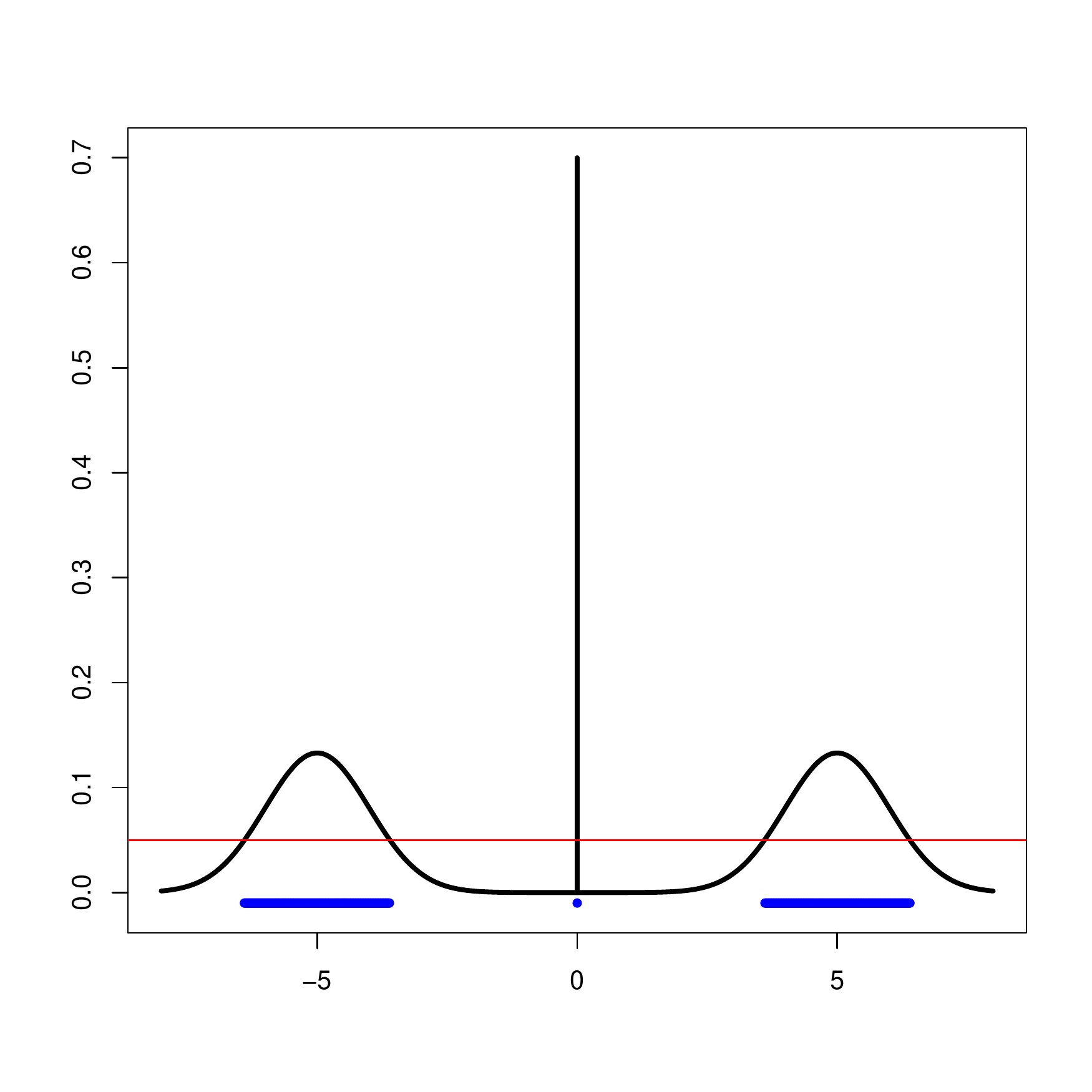} & 
\includegraphics[scale=.3]{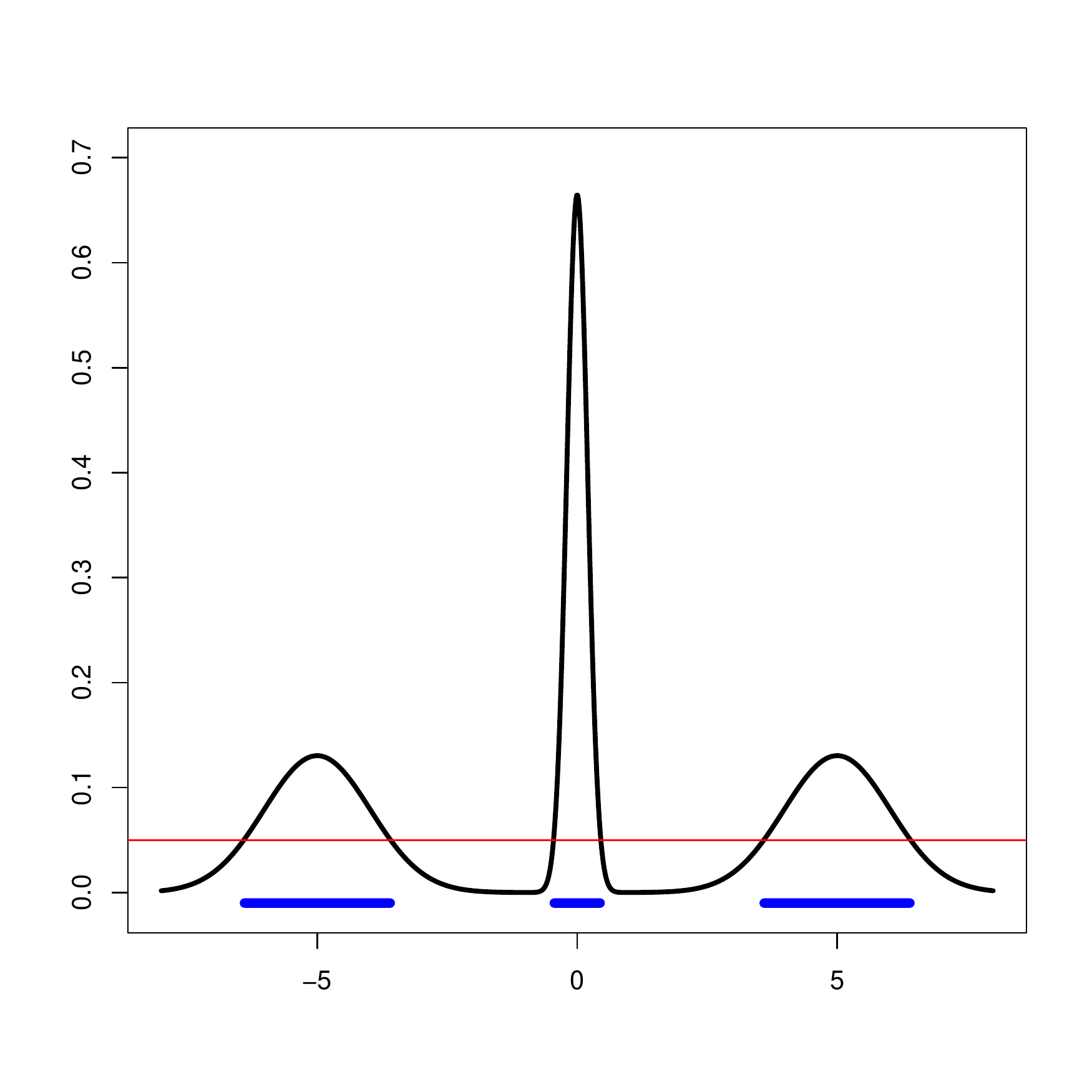}
\end{tabular}
\end{center}
\caption{Left: a distribution with a continuous component and a point mass at 0.
Right: the smoothed density $p_h$. The level set $L_h$ is biased but is estimable
and it approximates the main features of $L$.}
\label{fig::mix}
\end{figure}

As a second example,
let $P= (1/3) \phi(x;-5,1) + (1/3) \delta_0 + (1/3) \phi(x;5,1)$
where $\phi$ is a Normal density and $\delta_0$ is a point mass at 0.
Of course, this distribution does not even have a density.
The left plot in Figure
(\ref{fig::mix}) shows the density of the absolutely continuous
part of $P$ with a vertical line to who the point mass.
The right plot shows $p_h$, which is a smooth, well-defined density.
Again the blue points show the level sets.
As before $p_h$ is biased (as is $L_h$).
But $p_h$ is well-defined, as is $L_h$,
and $\hat p_h$ and $\hat L_h$
are accurate estimators of $p_h$ and $L_h$.
Moreover, $L_h$ contains the most important 
qualitative information about $L$, namely, that there are
three connected components, one of which is small.

The idea of viewing $p_h$
as the estimand is not new.
The ``scale space'' approach to smoothing
explictly argues that we should view
$\hat p_h$ as an estimate of $p_h$,
and $p_h$ is then regarded as a view of $p$ at a particular resolution.
This idea is discussed in detail in
\cite{chaudhuri2000scale, chaudhuri1999sizer, godtliebsen2002significance}.

If we do decide to base TDA on
tuning parameters that do not go to 0 as $n$ increases
then we need new methods for choosing tuning
parameters.
One possibility, suggested in
\cite{chazal2014robust} and
\cite{guibas2013witnessed}
is to choose the tuning parameter that maximizes the
number of significant topological features.
In particular, 
\cite{chazal2014robust} use the bootstrap to
assess the significance of topological features and then they choose
the smoothing parameter to maximize the number of such features.
This maximal significance approach is promising but so far
there is no theory to support the idea.

The problem of choosing tuning parameters thus remains
one of the greatest challenges in TDA.
In fact, the same problem permeates
the clustering literature.
To date, there is no agreement on how to choose $k$ in $k$-means
clustering, for example.

\section{DATA VISUALIZATION AND EMBEDDINGS}

Topological ideas play a role in data visualization
either explicitly or implicitly.
In fact, many TDA methods may be regarded as visualization methods.
For example, density trees, persistence diagrams and manifold learning
all provide low dimensional representations of the data
that are easy to visualize.

Some data visualization methods work by embedding the data
in $\mathbb{R}^2$ and then simply plotting the data.
Consider a point cloud $X_1,\ldots, X_n$ where
$X_i\in \mathbb{R}^d$.
Let $\psi:\mathbb{R}^d \to \mathbb{R}^2$ and let
$Z_i = \psi(X_i)$.
Because the points $Z_1,\ldots, Z_n$
are in $\mathbb{R}^2$,
we can easily plot the $Z_i$'s.
Perhaps the most familiar version is
multidimensional scaling (MDS)
where $\psi$ is chosen to be a linear function minimizing
some measure of distance between
the original pairwise distances $||X_i-X_j||^2$ and the
embedded distances
$||Z_i-Z_j||^2$.
In particular, if we minimize
$\sum_{i\neq j}(||X_i-X_j||^2 - ||Z_i-Z_j||^2)$
then the solution is to project the data onto the first two principal components.

But traditional MDS does a poor job of preserving local structure such as clusters.
Local, nonlinear versions of MDS do a better job of preserving local structure.
An example is
{\em Laplacian Eigenmaps}
which was proposed by
\cite{belkin2003laplacian}.
Here, we choose $\psi$ to minimize
$\sum_{i,j} W_{ij}||Z_i-Z_j||^2$
(subject to some consraints)
where the $W_{ij}$ are localization weights such as
$W_{ij} = e^{-||X_i-X_j||^2/(2h^2)}$.
The resulting embedding does a good job of preserving local structure.
However,
\cite{maaten2008visualizing} noted that local methods of this type can
cause the data to be too crowded together.
They proposed a new method called t-SNE which seems to work better but they provided
no justification for the method.
\cite{carreira2010elastic} provided an explanation of why t-SNE works.
He showed that t-SNE optimizes a criterion that essentially contains two terms,
one promoting localization and the other which causes points to repel each other.
Based on this insight,
he proposed a new method called {\em elastic embedding}
that explicitly has a term encouraging clusters to stay together and a term that repels
points from each other.
What is notable about t-SNE and elastic embedding is that they preserve clusters and loops.
The loops are preserved apparently due to the repelling term.
It appears, in other words that
these methods preserve topological features of the data.

This leads to the following question:
is it possible to derive low-dimensional embedding
methods that explicitly preserve topological features of the data?
This is an interesting open question.

\section{APPLICATIONS}
\label{section::applications}

\subsection{The Cosmic Web}

\begin{figure}
\begin{center}
\includegraphics[scale=.5]{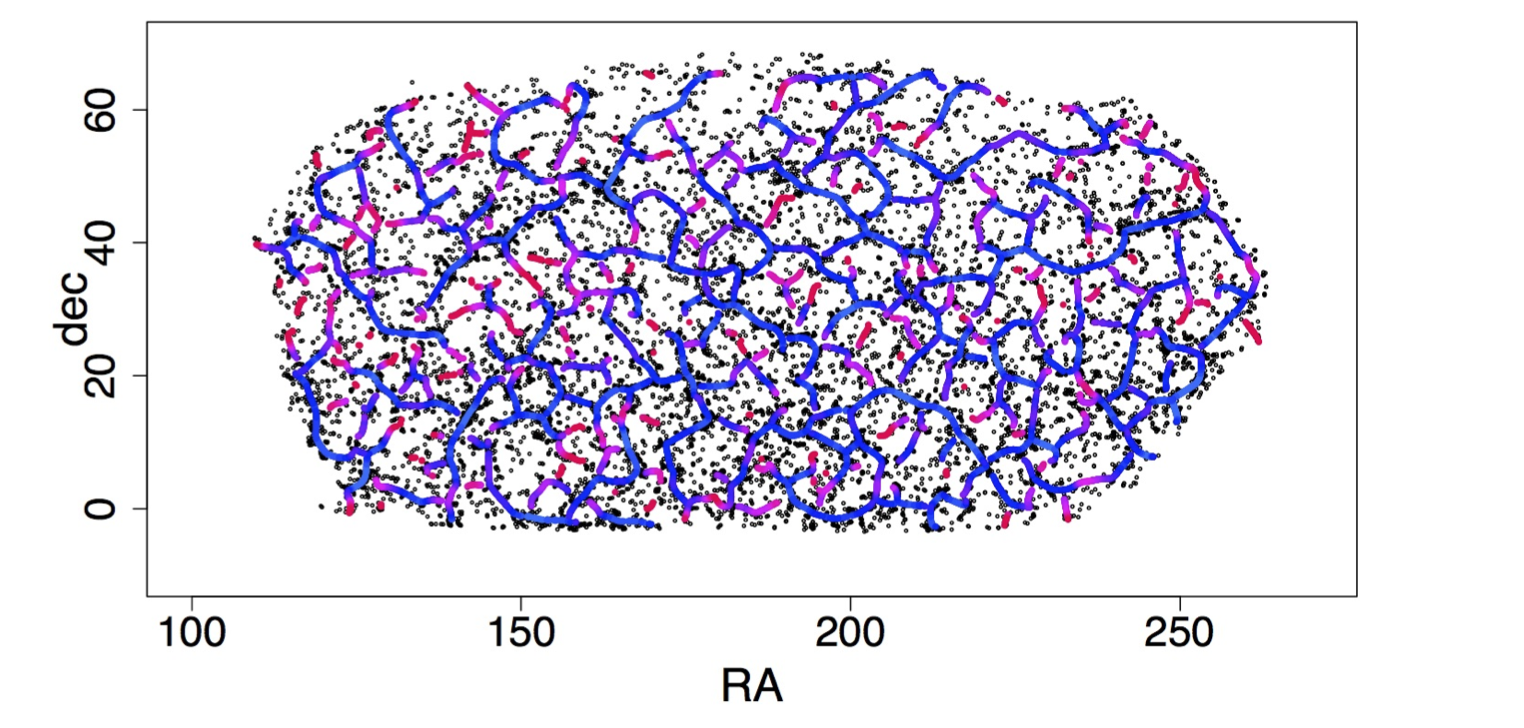}
\end{center}
\caption{
A filament map from \cite{chen2015cosmic}.
The data are galaxies from the Sloan Digital Sky Survey.
The blue lines are detected filaments.
The red dots are clusters.}
\label{fig::Cosmic-Web}
\end{figure}

The matter in the Universe is distributed in a complex, spiderweb-like pattern
known as the Cosmic web.
Understanding and quantifying this structure is one of the challenges of
modern cosmology.
Figure \ref{fig::Cosmic-Web}
shows a two-dimensional slice of data consisting of some galaxies
from the Sloan Digital Sky Survey
(\url{www.sdss.org})
as analyzed in \cite{chen2015cosmic}.
(RA refers to ``right ascension'' and DEC refers to ``declination.''
These measure position in the sky using essentially longitude and latitude).
The blue lines are filaments that were found using the
ridge methods discussed in Section \ref{section::ridges}.
Also shown are clusters (red dots) that were found by 
previous researchers.
Filament maps like this permit researchers
to investigate questions about how structure
formed in our Universe.
For example,
\cite{chen2015investigating}
investigated how the properties of galaxies
differ depending on the distance from filaments.

Several papers,
such as
\cite{van2011alpha,van2011probing,van2010alpha}
have used homology and persistent homology to
study the structure of the cosmic web.
These papers use TDA to quantify the clusters, holes and voids
in astronomical data.
\cite{sousbie2011persistent2,sousbie2011persistent}
uses Morse theory to model the filamentary structures of the cosmic web.

\subsection{Images}

\begin{figure}
\begin{center}
\includegraphics[scale=.3]{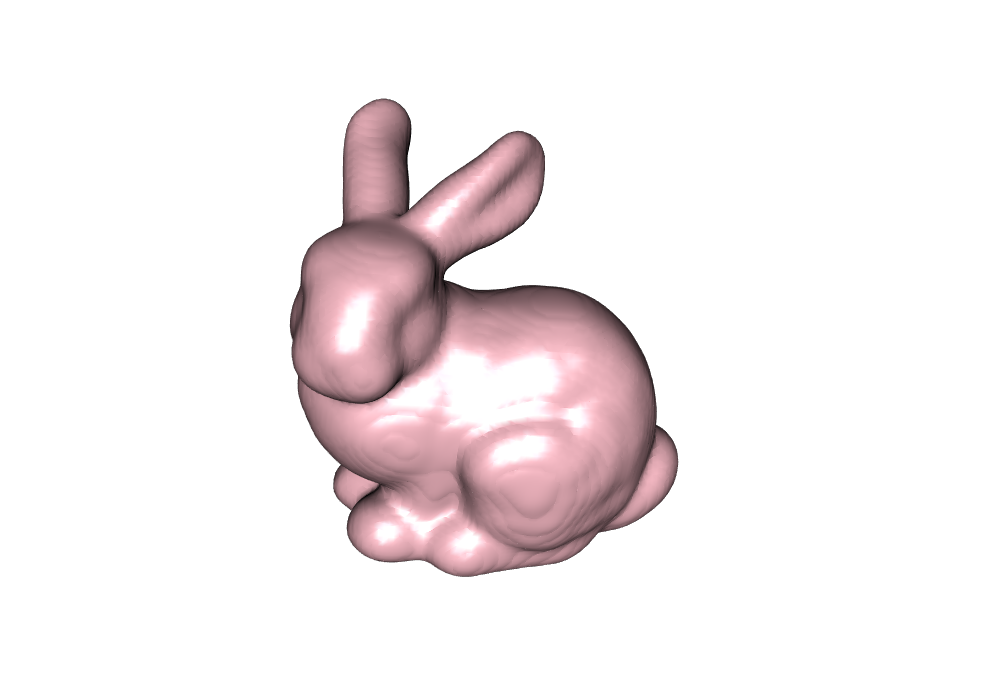}
\end{center}
\vspace{.2cm}
\caption{A three-dimensional image. Classifying such images requires features that
are invariant to small deformations of the image. TDA can potentially provide such features.}
\label{fig::rabbit}
\end{figure}

Many researchers have used some form of TDA
for image analysis.
Consider Figure \ref{fig::rabbit}
which shows a 3d image of a rabbit.
Given a large collection of such images, possibly corrupted by noise,
we would like to define features that can be used for classifying such images.
It is critical that the features be invariant
to shifts, rotations and small deformations.
Topological are thus a promising source of relevant features.
A number of papers have used TDA to
define such features, for example,
\cite{bonis2016persistence, li2014persistence,carriere2015stable}.

\begin{figure}
\begin{center}
\includegraphics[scale=.5]{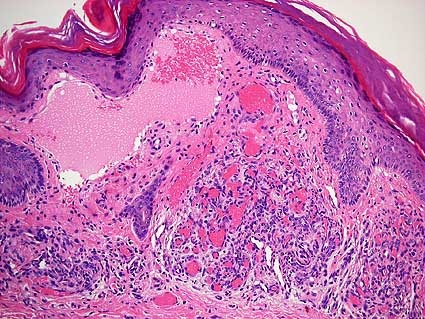}
\end{center}
\caption{An example of a histology image from
{\tt http://medicalpicturesinfo.com/histology/}.}
\label{fig::histology}
\end{figure}

TDA has also been used in the classification of 2d images.
For example
\cite{singh2014topological}
considered
breast cancer histology images.
These images show the arrangement of cells
of tissue samples.
An example of a histology image is given in Figure \ref{fig::histology}.

A typical image has many clumps and voids 
so TDA may be an appropriate method for summarizing the images.
\cite{singh2014topological}
used the Betti numbers as a function of scale,
as features for a classifier.
The goal was to discriminate different sub-types of cancer.
They achieved a classification accuracy of 69.86 percent.

\subsection{Proteins}

\begin{figure}
\begin{center}
\includegraphics[scale=.5]{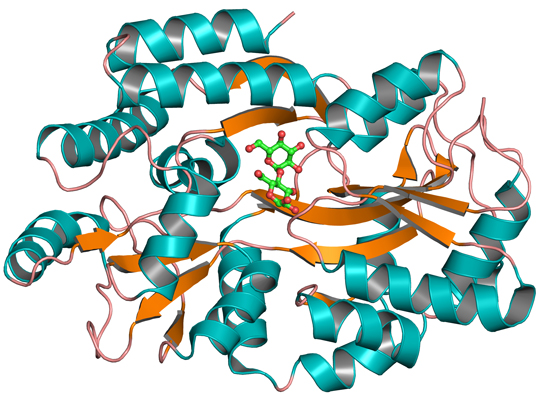}
\end{center}
\caption{A maltose binding protein.
The image of the protein is from 
\tt{http://lilith.nec.aps.anl.gov/Structures/Publications.htm}.}
\label{fig::protein}
\end{figure}

\cite{kovacev2016using}
used TDA to study
the maltose binding protein
which is a protein found in
Escherichia coli.
An example of such a protein is given in Figure 
\ref{fig::protein};
the figure is from \url{http://lilith.nec.aps.anl.gov/Structures/Publications.htm}.
The protein is a dynamic structure
and the changes in structure are of biological relevance.
Quoting from \cite{kovacev2016using}:
\begin{quote}
A major conformational change in the protein occurs when a smaller
molecule called a ligand attaches to the protein molecule ...
Ligand-induced
conformational changes are important because the biological function
of the protein occurs through a transition from a ligand-free (apo) to
a ligand-bound (holo) structure ...
\end{quote}
The protein can be in an open or closed conformation,
and the closed conformation is due to having a captured ligand.
The goal of the authors is to classify the state of the protein.

Each protein is represented by 370 points 
(corresponding to amino acids) in three dimension space.
The authors
construct a
dynamic model of the protein structure 
(since the structure changes over time)
from which they define
dynamical distances between the 370 points.
Thus a protein is represented by a 370 by 370 distance matrix.
From the distance matrix they construct a persistence diagram.
Next, they convert the persistence diagrams
into a set of functions called
{\em landscapes} as defined in
\cite{bubenik2015statistical}.
Turning the diagram into a set of one-dimensional functions
makes it easier to use standard statistical tools.
In particular, they do a two-sample permutation test
using the integrated distances between the landscape functions
as a test statistic.
The p-value is $5.83\times 10^{-4}$
suggesting a difference between the open and closed conformations.
This suggests that landscapes can be used to classify proteins as open or closed.
They also show that certain sites on the protein, known as {\em active sites},
are associated with loops in the protein.

\subsection{Other Applications}

Here I briefly mention
a few other examples of TDA.

The Euler characteristic is a topological quantity which I did not
mention in this paper. It has played an important role
in various aspects of probability as well as to applications in astrophysics and neuroscience
\citep{worsley1995boundary,taylor2007detecting,adler2009random,worsley1994local,worsley1996geometry, taylor2007detecting}.
The Euler characteristic has also been used for
classification of shapes \citep{richardson2014efficient}.
See also \cite{turner2014persistent}.
\cite{bendich2010computing}
use topological methods to
study the interactions between root systems of plants.
\cite{carstens2013persistent} use persistent homology to describe the
structure of collaboration networks.
\cite{xia2015multiresolution}
use TDA 
in the analysis of biomolecules.
\cite{adcock2014classification}
use TDA to classify images of lesions of the liver.
\cite{chung2009persistence}
use persistence diagrams constructed from data
on cortical thickness
to distinguish control subjects and austistic subjects.
\cite{offroy2016topological}
reviews the role of TDA in chemometrics.
\cite{bendich2016persistent} use persistent homology to study the structure of brain arteries.
There is now a substantial literature on TDA in neuroscience
including
\cite{arai2014effects,
babichev2016persistent,
basso2016gamma,
bendich2014persistent,
brown2012structure,
cassidy2015brain,
chen2014neural,
choi2014abnormal,
chung2009persistence,
curto2008cell,
curto2013neural,
curto2015makes,
curto2015neural,
curto2016what,
dabaghian2011topological,
dabaghian2012topological,
dabaghian2014reconceiving,
dabaghian2015geometry,
dlotko2016topological,
ellis2014describing,
giusti2013no,
giusti2015clique,
giusti2016twos,
hoffman2016topological,
jeffs2015sparse,
kanari2016quantifying,
khalid2014tracing,
kim2014morphological,
lee2011discriminative,
lienkaemper2015obstructions,
manin2015neural,
masulli2015topology,
petri2014homological,
pirino2014topological,
singh2008topological,
sizemore2016classification,
sizemore2016closures,
spreemann2015using,
stolz2014computational,
yoo2016topological,
zeeman1962topology}.
The website
\url{http://www.chadgiusti.com/algtop-neuro-bibliography.html}
maintain a bibliography of references in this area.

\section{CONCLUSION: THE FUTURE OF TDA}

TDA is an exciting area and is full of interesting ideas.
But so far, it has had little impact on data analysis.
Is this because the techniques are new?
Is it because the techniques are too complicated?
Or is it because the methods are simply not that useful in practice?

Right now, it is hard to know the answer.
My personal opinion is that TDA is 
a very specialized tool that is useful
in a small set of problems.
For example, it seems to be an excellent tool for
summarizing data relating to the cosmic web.
But, I doubt that TDA will ever become
a general purpose tool like regression.
The exception is clustering, which of course is used routinely,
although some might argue that it is a stretch to consider clustering part of TDA.
I have seen a number of examples where complicated TDA methods were used to
analyze data but no effort was made to compare these methods to simpler,
more traditional statistical methods.
It is my hope that, in the next few years,
researchers will do thorough comparisons of standard statistical methods
with TDA in a number of scientific areas so that we can truly assess the value of these
new methods.

\section*{DISCLOSURE STATEMENT}
The author is not aware of any affiliations, memberships,
funding, or financial holdings that might be perceived as affecting
the objectivity of this review.

\section*{ACKNOWLEDGMENTS}

Thanks to Steve Fienberg for
suggesting that I write this review.
And thanks to 
Robert Adler,
Eddie Amari, 
Omer Bobrowski,
Bertrand Michel,
Vic Patrangenaru,
JaeHyeok Shin, 
Isabella Verdinelli, 
for providing comments
on an earlier draft.


\bibliographystyle{ar-style1}
\bibliography{paper}

\end{document}